\documentclass[sigconf]{acmart}

\usepackage{color}
\usepackage{booktabs} 
\usepackage{tabularx}		
\usepackage[table]{}
\usepackage{amsmath,graphicx,subcaption,balance}
\usepackage[english]{babel}
\usepackage[utf8]{inputenc}
\usepackage[colorinlistoftodos]{todonotes}
\usepackage[hyphens]{} 
\usepackage{MnSymbol} 		
\usepackage{colortbl}
\usepackage{hhline}

\usepackage{algorithm}
\usepackage[noend]{algpseudocode}

\usepackage{enumitem}   
\setlist{noitemsep}
\setlist{nolistsep}

\graphicspath{{./figures/}}
\usepackage{numprint}
\usepackage{multirow}

\newcommand{\hide}[1]{}    

\usepackage{mathtools}
\newcommand{\define}{\coloneqq}							

\usepackage{subcaption} 	
\usepackage{xspace}			
\newcommand{\candedge}{\texttt{CandEdge}\xspace}
\newcommand{\candnode}{\texttt{CandNode}\xspace}
\newcommand{\graphedge}{\texttt{GraphEdge}}

\newcommand{\introparagraph}[1]{\textbf{#1.}}        

\newcommand{\cc}{gray}
\newcommand{\algocomment}[1]{\textcolor{\cc}{ #1}}

\newcommand{\eg}{{\it e.g.,\ }}
\newcommand{\etal}{{\it et al.\ }}

\newcommand{\ouralgorithmname}{KARPET}
\newcommand{\baselineone}{Unguided}
\newcommand{\baselinetwo}{Backbone}
\newcommand{\Vertices}{V}
\newcommand{\Edges}{E}
\newcommand{\LabelingFunct}{\varphi}
\newcommand{\LabelSet}{L}
\newcommand{\WeightFunct}{w}
\newcommand{\Graph}{G}
\newcommand{\TreePattern}{Q}
\newcommand{\VerticesPattern}{V_Q}
\newcommand{\EdgesPattern}{E_Q}
\newcommand{\TreeRoot}{\top}
\newcommand{\LabelConstraint}{\psi}

\newcommand{\Reals}{\mathbb{R}}

\newcommand{\Terminals}{\bot}

\newcommand{\Neighbours}{N}
\newcommand{\BijectionFunct}{\lambda}

\newcommand{\Priority}{\textsc{Weight}\xspace}
\newcommand{\Candidates}{\mathrm{Candidates}}

\begin{document}

\copyrightyear{2018}
\acmYear{2018} 
\setcopyright{iw3c2w3}
\acmConference[WWW 2018]{The 2018 Web Conference}{April 23--27, 2018}{Lyon, France}
\acmBooktitle{WWW 2018: The 2018 Web Conference, April 23--27, 2018, Lyon, France}
\acmPrice{}
\acmDOI{10.1145/3178876.3186115}
\acmISBN{978-1-4503-5639-8/18/04}

\fancyhead{}

\title{Any-k: Anytime Top-k Tree Pattern Retrieval in Labeled Graphs}

\author{Xiaofeng Yang}
\affiliation{
       \institution{Northeastern University}
       \streetaddress{360 Huntington Ave.}
       \city{Boston}
       \state{MA}
       }
 \email{xiaofeng@ccs.neu.edu}
 
\author{Deepak Ajwani}
\affiliation{
       \institution{Nokia Bell Labs}
       \city{Dublin}
       \state{Ireland}
       }
 \email{deepak.ajwani@nokia-bell-labs.com}
 
\author{Wolfgang Gatterbauer}
\affiliation{
       \institution{Northeastern University}
       \streetaddress{360 Huntington Ave.}
       \city{Boston}
       \state{MA}
       }
 \email{wolfgang@ccs.neu.edu}
 
\author{Patrick K. Nicholson}
\affiliation{
       \institution{Nokia Bell Labs}
       \city{Dublin}
       \state{Ireland}
       }
 \email{pat.nicholson@nokia-bell-labs.com}

 \author{Mirek Riedewald}
\affiliation{
       \institution{Northeastern University}
       \streetaddress{360 Huntington Ave.}
       \city{Boston}
       \state{MA}
       }
 \email{mirek@ccs.neu.edu}

 \author{Alessandra Sala}
\affiliation{
       \institution{Nokia Bell Labs}
       \city{Dublin}
       \state{Ireland}
       }
\email{alessandra.sala@nokia-bell-labs.com}

\renewcommand{\shortauthors}{X. Yang et al.}

\begin{abstract}
Many problems in areas as diverse as recommendation systems, social network analysis, semantic search, and distributed root cause analysis can be modeled as pattern search on labeled graphs (also called ``heterogeneous information networks'' or HINs). 
Given a large graph and a query pattern with node and edge label constraints, a fundamental challenge is to find the top-$k$ matches according to a ranking function over edge and node weights. 
For users, it is difficult to select value $k$.
We therefore propose the novel notion of an \emph{any-$k$ ranking algorithm}: 
for a given time budget, return as many of the top-ranked results as possible. 
Then, given additional time, produce the next lower-ranked results quickly as well.
It can be stopped anytime, but may have to continue until all results are returned. 
This paper focuses on acyclic patterns over arbitrary labeled graphs. We are interested in practical algorithms that effectively exploit (1) properties of heterogeneous networks, in particular selective constraints on labels, and 
(2) that the users often explore only a fraction of the top-ranked results. 
Our solution, {\ouralgorithmname}, carefully integrates aggressive pruning that leverages the acyclic nature of the query, and incremental guided search. 
It enables us to prove strong non-trivial time and space guarantees, which is generally considered very hard for this type of graph search problem. 
Through experimental studies we show that {\ouralgorithmname} achieves running times in the order of milliseconds for tree patterns on large networks with millions of nodes and edges.
\end{abstract}

\maketitle

\section{Introduction}
\label{sec:intro}

\begin{figure}[tb]
    \centering
    \includegraphics[width=0.7\linewidth]{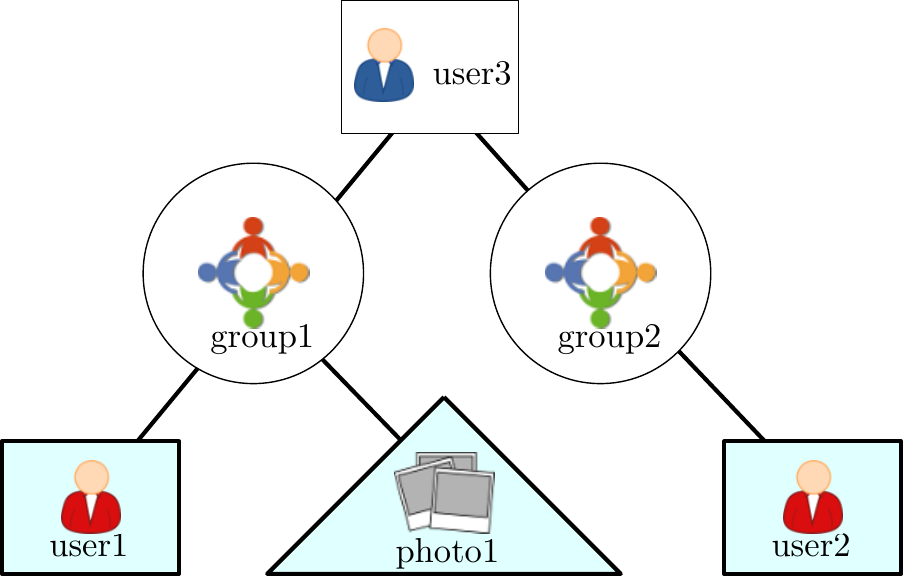}
    \caption{Example query on a photo-sharing network: 
	find the most important nodes of types ({user3}, {group1}, {group2}) 
	for a given triple of specified nodes of types ({photo1}, {user1}, {user2}).}
    \label{fig:FlickrExample}
\end{figure}

Heterogeneous information networks (HIN)~\cite{han2010mining}, i.e., graphs with node and/or edge labels, have recently attracted a lot of attention for their ability to model many complex real-world relationships, thereby enabling rich queries. 
Often labels are used to represent types of nodes and their relationships:

\begin{example}[Photo-sharing network]\label{ex:flickr1}
Consider a photo-sharing social network with three vertex type labels: user, photo, and group. Users are connected to the photos they upload, and photos are connected to groups when they are posted there. Finally, users can connect to groups by joining them. To maintain a vibrant community and alert users about potentially interesting photos, the social network might run queries of the type shown in Figure~\ref{fig:FlickrExample}: 
given \textit{photo1} and two users, \textit{user1} and \textit{user2}, find alternative groups (matching nodes for \textit{group2}) 
to post the photo in order to reach \textit{user2} without spamming her directly. 
This is achieved by identifying a user belonging to both groups (\textit{user3}), who can post the photo in the other group. 
There might be hundreds of matching triples (\textit{group1}, \textit{user3}, \textit{group2}), and there would be many more if \textit{user2} was not given in advance. 
Under these circumstances, the goal often is not to find \emph{all} results, but only the \emph{most important} ones. 
Importance can be determined based on node and edge weights, e.g., weights representing distances (or similarities). 
Then the query should return the lightest (or heaviest) pattern instances. 
For example, the weight of a group may be based on its number of members, the weight of a user on how active s/he is, and the weight of a link on the timestamp when it was established (to give preference to long-term relationships or more recent photo posts), or the sum of the PageRanks of its endpoints.
\end{example}

These types of rich query semantics also appears in other contexts, e.g., root-cause analysis in distributed systems. 
The Vitrage service for OpenStack~\cite{Vitrage}
makes use of path and tree patterns to specify rules for automatic root cause deduction of alarms raised by virtual machines and hardware.
Large OpenStack deployments---involving thousands of hosts and tens of thousands of virtual machines and hardware components---necessitate pattern matching algorithms to deduce the root cause of such patterns in near real-time.

We focus on efficient solutions for acyclic pattern queries on general labeled graphs. 
To this end, we propose the notion of \emph{any-$k$} algorithms, a novel variant of top-$k$ algorithms. A top-$k$ algorithm exploits knowledge about the given $k$ to produce the top-$k$ lightest patterns faster than the ``full enumeration'' algorithm 
(which first produces all results and then ranks them by weight). 
In practice, it is difficult for users to know the value of $k$ upfront (``when will I have seen enough?''). An any-$k$ algorithm addresses this issue by not requiring a pre-set value for $k$. Instead, an any-$k$ algorithm
\begin{enumerate}
\item returns the top-ranked result as quickly as possible,
\item then returns the second-ranked result next, followed by the third-ranked, and so on,
\item until the user is satisfied and terminates the process.
\end{enumerate}
In other words, the ranked enumeration can be stopped \emph{anytime} and should then return as many top results as possible.

The queries we are interested in correspond to \emph{subgraph isomorphism}, which is known to be hard in general.
In particular, subgraph isomorphism on homogeneous graphs is 
already NP-complete in the size of the query
(even for the path case as Hamiltonian path is a special case). And labeled graphs contain unlabeled graphs as a special case. 
On the other hand, labels provide more opportunities for achieving better performance \emph{in practice} by exploiting heterogeneity where present. Note that a key reason for hardness of isomorphism lies in the ``non-repetition constraint,'' i.e., the same graph node cannot occur more than once in an answer. Without this constraint, pattern search would correspond to the easier \emph{subgraph homomorphism} problem
which can be solved in PTIME.

Our approach is based on three key insights:
(1) Constraints on node or edge labels can dramatically reduce the number of matching results; 
(2) Mutually exclusive {type} labels ``narrow the gap'' in cardinality between the set of isomorphic subgraphs and the set of homomorphic subgraphs (which includes all isomorphic ones). 
The reason is that query pattern nodes of different types cannot be mapped to the same graph node, even when the algorithm is only searching for homomorphism. 
In the example photo-sharing network, users and photos cannot stand in for a group node. In the extreme, if all nodes in the query pattern have different types, then any solution for subgraph homomorphism also satisfies isomorphism. This suggests an approach that aggressively prunes for the homomorphism case and then filters based on node repetitions in the result patterns; and 
(3) In many real-world cases, output size is small relative to the combinatorial size of the pattern search space. Hence algorithm complexity bounds based on output size promise to deliver practically meaningful performance guarantees.

\textbf{Overview of the Solution.} Our approach combines three conceptually separate steps into a two-phase algorithm.

1) The search space of possible homomorphic patterns is pruned to the provably smallest representation of the original graph. We use insights from the well-known Yannakakis algorithm~\cite{DBLP:conf/vldb/Yannakakis81} for evaluating answers to acyclic conjunctive queries to create this representation in just one bottom-up and a subsequent top-down sweep through the query tree.

2) We devise a novel any-$k$ algorithm for enumerating homomorphic tree patterns. It uses dynamic programming to perform a bottom-up cost calculation, followed by a top-down guided search.

3) A final pruning step removes those homomorphic patterns that do not satisfy the isomorphism requirement.

We show how to combine the first two steps into just one bottom-up and one top-down phase.
We then integrate the third step into the combined top-down phase. Our experiments show that even on graphs with millions of nodes and tens of millions of edges, we can return the top-ranked results in just a few milliseconds, whereas alternative approaches would take orders of magnitude longer. Our implementation can be downloaded from~\cite{Code2018}.

\textbf{Main contributions.} We devise {\ouralgorithmname}  
(\underline{K}ernelization\footnote{\emph{Kernelization} is a pre-processing technique that replaces the original input by a (usually) smaller representation called ``kernel'' in order to reduce the computation cost. Our approach enumerates solutions over a smaller pruned candidate graph.}
\underline{A}nd \underline{R}apid \underline{P}runing-based \underline{E}xploration for \underline{T}ree patterns), a novel and highly performant any-$k$ algorithm that can quickly identify top-ranked tree patterns in large graphs, then return the next lower-ranked ones when given extra time. 

1) {\ouralgorithmname} is designed as an \emph{anytime} ranking algorithm that enumerates homomorphic subtrees in order of total edge weight with strong theoretic guarantees: We show that our worst-case time complexity for returning all homomorphism results is identical to full enumeration.
In addition, {\ouralgorithmname} provides strong upper bound guarantees for the time to return the top-ranked homomorphism result, as well as the time between returning a homomorphism result and the next. For cases with ``small gap'' between homomorphism and isomorphism, i.e., when ``sufficiently many'' homomorphic patterns are also isomorphic patterns, these guarantees carry over to subgraph isomorphism.

2) We propose fast and effective \emph{local} pruning operations that exploit the heterogeneity of labeled graphs, proving that they also guarantee strong \emph{global} pruning properties. Intuitively, for subgraph homomorphism, we show that inexpensive pruning based on 1-node neighborhoods efficiently removes all candidate nodes that are not part of any result pattern.

3) In contrast to a lot of theoretical work on subgraph isomorphism algorithms, our algorithm is output-sensitive---its worst case complexity depends on the output size, which is smaller when the graph and the query are more heterogeneous, rather than being exponential in the size of the query pattern.

4) We show how to speed up the search for top-ranked isomorphic answers by pushing the pruning for non-repeating nodes into the incremental result enumeration algorithm.

\section{Problem Definition and Hardness}
\label{sec:problem}

\emph{Our goal is to find the lightest subgraphs of a labeled graph $\Graph$ that are isomorphic to a given tree pattern $\TreePattern$}. Instead of returning all results at once after a long wait time, we set out to devise an \emph{anytime algorithm}, which returns the top-ranked match as quickly as possible and then incrementally returns the remaining results over time.

\begin{definition}[Any-$k$ algorithm]
	An any-$k$ algorithm is a variant of a top-$k$ algorithm in which $k$ is not known at the start of the algorithm. 
	The algorithm can be interrupted \emph{anytime}, returning the top-$k$ results with $k$ being as large as possible.
\end{definition}	

We define the weight of a pattern as the \emph{sum} of edge weights.
This also supports search for the ``most reliable'' pattern based on probabilities assigned to edges. 
Finding the pattern with the greatest probability of being connected, assuming independence, is equivalent to maximizing the sum of the logarithms of the edge probabilities.
For our problem with a fixed query pattern, lightest and heaviest pattern search can be easily converted into each other. It is also straightforward to modify our approach to support pattern weight defined as minimum or maximum of edge weights.
We present the formal definitions next. Table~\ref{tab:notation} summarizes important notation.

\begin{table}[t]
\caption{Notation used in this paper}\label{tab:notation}
\centering
\small
\begin{tabularx}{\linewidth}{  @{\hspace{0pt}} >{$}l<{$}  @{\hspace{2mm}}  X @{}}
\hline
\textrm{Symbol}			& Definition 	\\
\hline
\Graph(\Vertices, \Edges)		& A labeled graph with node set $\Vertices$ and edge set $\Edges$ \\
\LabelSet		        & Set of node labels \\
\LabelingFunct()   & Function mapping nodes to labels \\
\WeightFunct()     & Function mapping edges to weights \\
\TreePattern(\VerticesPattern, \EdgesPattern)     & Tree pattern with node set $\VerticesPattern$ and edge set $\EdgesPattern$ \\
\LabelConstraint()   & Required labels for a graph node matched to a query node \\
\Terminals(\TreePattern)       & Set of leaf nodes (or terminals) in $\TreePattern$ \\
\TreeRoot(\TreePattern)       & Chosen root node in $\TreePattern$ \\
\Neighbours(v,\ell)      & Set of neighbors of $v$ in $\Graph$ with label $\ell$ \\
\BijectionFunct() & Function mapping query nodes $V_Q$ to graph nodes $V$ \\
\hline
\end{tabularx}
\end{table}

\begin{definition}[HIN, labeled graph]
A \emph{Weighted Heterogeneous Information Network} (HIN) is a 
labeled undirected graph $\Graph = (\Vertices,\Edges,\LabelingFunct, \WeightFunct)$, 
where $\Vertices$ is a set of vertices, $\Edges$ is a set of edges,  
$\LabelingFunct$ is a node labeling function $\LabelingFunct : \Vertices \rightarrow \LabelSet$,
and $\WeightFunct$ is an edge weight function $\WeightFunct : \Edges \rightarrow \Reals$.
\end{definition}

In many HINs, a node has at least two different kinds of labels: a unique \emph{node ID} and a \emph{type} (or \emph{class}). In the photo-sharing network example (see Example~\ref{ex:flickr1}), the labeling function assigns types such as ``user'' or ``photo'' to each node.
Our approach can be easily extended to include multiple labels per node, as well as (multiple) edge labels, node weights, and directed edges. We omit these straightforward generalizations in order to simplify the exposition. 

Given a vertex $v \in \Vertices$ and label $\ell \in \LabelSet$, we use $\Neighbours(v, \ell)$ to denote the set of all \emph{neighbors} of $v$ with label $\ell$, i.e., $\Neighbours(v, \ell) \define \{u : (v, u) \in \Edges \land \LabelingFunct(u) = \ell \}$.

\begin{definition}[Tree pattern or query $Q$]
Given a labeled graph $\Graph = (\Vertices,\Edges,\LabelingFunct, \WeightFunct)$, 
a \emph{tree pattern} is a rooted tree 
$\TreePattern = (\VerticesPattern,\EdgesPattern, \LabelConstraint)$
in which each node $v \in \VerticesPattern$ has a label constraint 
$\LabelConstraint : \VerticesPattern \rightarrow \LabelSet$.
We use $\TreeRoot(\TreePattern) \in \VerticesPattern$
to denote the root of the tree and $\Terminals(\TreePattern)$ 
to denote the set of its leaves (or terminals, i.e.\ nodes of degree one).
\end{definition}

The labeling constraint can encode the selection of specific nodes or node types.
For example, in the photo-sharing network scenario, setting $\LabelConstraint$ for \textit{user1} to be the ID of a specific user node limits the candidate set for \textit{user1} to just this one graph node. Similarly, setting $\LabelConstraint$ to the label encoding the type ``group'' will enforce that only graph nodes representing groups, but not users or photos, will be considered.

Notice that $\TreePattern$ being rooted is not a restriction: any node in a tree can be chosen to be the root.
We merely make use of the fact that the tree pattern is rooted in order to more easily describe our algorithms.

\begin{definition}[homomorphic match or result pattern]
A \emph{homomorphic result pattern} (or \emph{homomorphic match}) of query $\TreePattern$ is a graph $(\Vertices' \subseteq \Vertices,\Edges' \subseteq \Edges)$ such that there exists a function $\BijectionFunct: \VerticesPattern \rightarrow \Vertices'$ with the following properties:
(1) $\forall u \in \VerticesPattern: \LabelConstraint(u) = \LabelingFunct(\BijectionFunct(u))$, and 
(2) $\forall (u,v) \in \EdgesPattern: (\BijectionFunct(u), \BijectionFunct(v)) \in \Edges'$. 
The weight of a result pattern is defined as $\sum_{(u,v) \in \Edges'}\WeightFunct(u,v)$.
\end{definition}

\begin{definition}[match or result pattern]
An (isomorphic) \emph{result pattern} (or \emph{match}) of query $\TreePattern$ is a homomorphic result pattern $(\Vertices' \subseteq \Vertices,\Edges' \subseteq \Edges)$ with a bijective mapping function
$\BijectionFunct: \VerticesPattern \xrightarrow[]{\text{1:1}} \Vertices'$.
\end{definition}

The above definitions make it clear that the set of isomorphic matches is a subset of the homomorphic matches; and can be obtained by removing all those homomorphic matches where multiple query nodes are mapped to the same graph node.

In the discussion below we will also refer to \emph{partial patterns} 
(or \emph{partial matches}) for intermediate results of the computation. 
These are incomplete instances where some of the query nodes are mapped to NIL by $\BijectionFunct$. The \emph{direct successor} of a partial match is one where exactly one of the NIL targets is replaced by a graph node, growing the pattern by one additional node. With \emph{successor} we refer to any partial or complete match in the transitive closure of direct successor.

For fast access to $\Neighbours(v, \ell)$, we rely on \graphedge, a \emph{two-level hash index} constructed offline for $\Graph$. It maps a given node ID $v$ to another hash table, which in turn maps a given label $\ell$ to the set $N(v, \ell)$ of neighbors of $v$ with label $\ell$.
If no label is specified, all nodes and corresponding edge weights in the secondary hash table for $v$ are returned.
This index can be bulk-created from scratch in time linear in the graph size,
and updated in time linear in the size of the changes.

\textbf{Hardness.} In general, even the decision version of sub-graph isomorphism, i.e., to determine if a given query graph is isomorphic to a sub-graph of $\Graph$, is NP-complete. When the sub-graph is connected acyclic (i.e., a tree), the best worst-case time bound for the decision problem is a parameterized algorithm of Koutis and Williams~\cite{KW16} that requires $O(2^{|\Vertices_\TreePattern|}\texttt{poly}(|\Vertices|))$ time. Their algorithm also has matching conditional lower bounds~\cite{KT17}: achieving a bound of $O(2^{(1-\varepsilon)|\Vertices_\TreePattern|}\texttt{poly}(|\Vertices|))$ time, for any constant $\varepsilon>0$, would falsify a longstanding conjecture. Note that, since the decision problem is hard, the any-$k$ problem discussed here is at least as hard.

\begin{figure}[tb]
    \centering
    \includegraphics[width=0.95\linewidth]{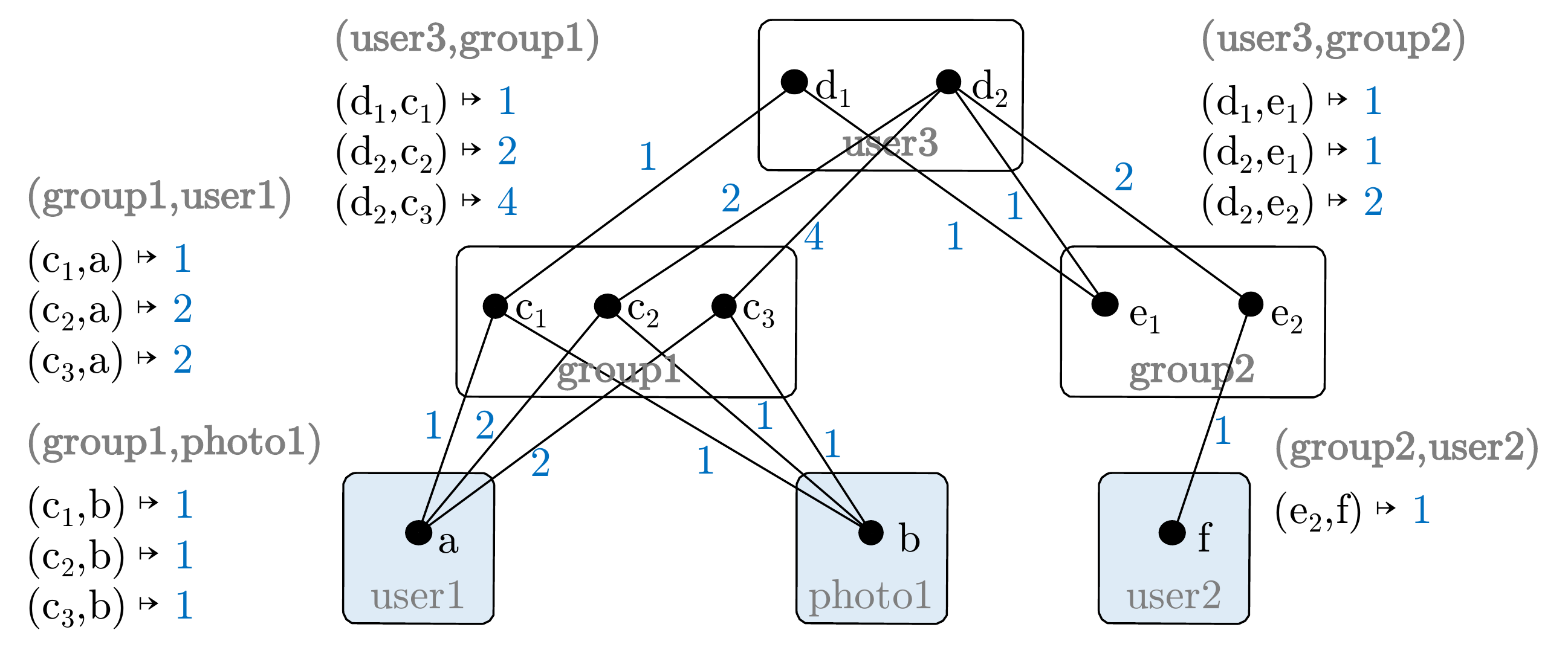}	
    \caption{Candidate instances for matching the example query in Figure~\ref{fig:FlickrExample}. 
	Edge sets are named based on the corresponding pairs of adjacent nodes in the query pattern.}
    \label{fig:joinGraph}
\end{figure}

In practice we often know specific node instances such as $a$ for \textit{user1} in the photo-sharing network example, and can dramatically reduce the pattern search space by exploring $\Graph$ starting from these nodes. Still, as \autoref{fig:joinGraph} illustrates, one cannot tell from the immediate neighborhood of node $a$, if edge $(a, c_1)$ will belong to top-ranked results, or any results at all. Worse yet, not even the 3-hop neighborhood of $a$ will answer this question. Hence a pattern search algorithm might suffer from expensive backtracking or the inability to determine, without extensive graph traversal, when the top-$k$ lowest-weight patterns have been found.

To the best of our knowledge, {\ouralgorithmname} is the first algorithm for ranked retrieval of graph query patterns that performs pruning and exploration based on ``local'' information, while \emph{provably guaranteeing} to make the right decisions ``globally.''

\section{Any-k Algorithm}
\label{sec:algorithm}

We next present an approach for sub-graph \emph{homomorphism}; this is a relaxation of sub-graph isomorphism in that
we do not require the mapping $\BijectionFunct$ from query nodes to tree-pattern nodes to be bijective (in other words, a node can be repeated in the result pattern).
Section~\ref{sec:repetitions} extends the approach for isomorphism.

{\ouralgorithmname} consists of two phases:
1) a bottom-up sweep from leaves to the root of $\TreePattern$, and 
2) a top-down depth-first traversal from root to leaves.
The first phase prunes some of the spurious candidates and creates a ``\emph{candidate graph}'' (discussed below) with ``\emph{minimum subtree weights}.'' The second phase prunes the remaining spurious candidates and performs a search guided by the subtree weights. Here the term \emph{spurious candidate} refers to a node or edge of the input graph that does not appear in any of the query results.

\subsection{Bottom-Up Phase}
\label{sec:bottomUp}

\setlength{\textfloatsep}{0pt}
\begin{algorithm}
  \small
  \caption{\label{alg:pruning}Bottom-up Subtree Weight Computation}
\begin{flushleft}  
\textbf{Input}: query $\TreePattern$, node neighborhood index $N(v, \ell)$ \\
\textbf{Output}: $\candnode: u \mapsto [c \mapsto [u' \mapsto w_{\min}]]$ \\
\hspace{0mm}\protect{\phantom{\textbf{Output}:}} $\candedge: (u, u') \mapsto [c \mapsto c']$	\\
\end{flushleft}
\begin{algorithmic}[1]

\State \algocomment{// For each leaf node in the query tree, 
find graph nodes with required label, add them to the candidates, and set their weights to 0}
\For{$u \in \Terminals(\TreePattern)$}	\label{alg:line:terminal}
	\State $\forall c \in \Vertices. 
	\LabelingFunct(c) = \LabelConstraint(u) :
	\candnode(u).\textsc{Insert}(c \mapsto (\textrm{NIL} \mapsto 0))$
\EndFor

\State \algocomment{// Traverse remaining query nodes in any bottom-up order}
\For{$u \in \textsc{Traversal}(\VerticesPattern)$}	\label{alg:line:traversal}

    \State \algocomment{//($i$) Find candidate edges adjacent to candidates in all children $u'$}
    \For{children $u'$ of $u$ in $\TreePattern$, and candidates $c' \in \candnode(u')$}
        \For{neighbors $c \in \Neighbours(c',\LabelConstraint(u))$} \label{alg:line:createcandidates}
          \State $\candedge(u, u').\textsc{Insert}(c \mapsto c')$
        \EndFor
    \EndFor
	
	\State \algocomment{//($\textit{ii}$) Keep only candidates with edges to each of the children of $u$}
	\State $C = \bigcap_{u' \textrm{ child nodes of } u} \candedge(u, u').\textsc{Keys}$	\label{alg:line:intersection}

	\State \algocomment{//($\textit{iii}$) Find min subtree weights for reachable candidates to children}
	\For{$c \in C$ and all children $u_i$ of $u$}
		\State $C' = \candedge(u, u').\texttt{Get}(c)$	
		\State $w_{i} \leftarrow \min_{c' \in C'}[w(c, c') + \Priority(c')]$		
		\State $\candnode(u).\textsc{Insert}(c \mapsto (u_i \mapsto w_i))$	\label{alg:line:minsubtree}
	\EndFor
	  
\EndFor

\end{algorithmic}
\end{algorithm}

\begin{figure*}[t]
\centering

\begin{subfigure}[t]{.31\linewidth}
	\includegraphics[scale=0.31]{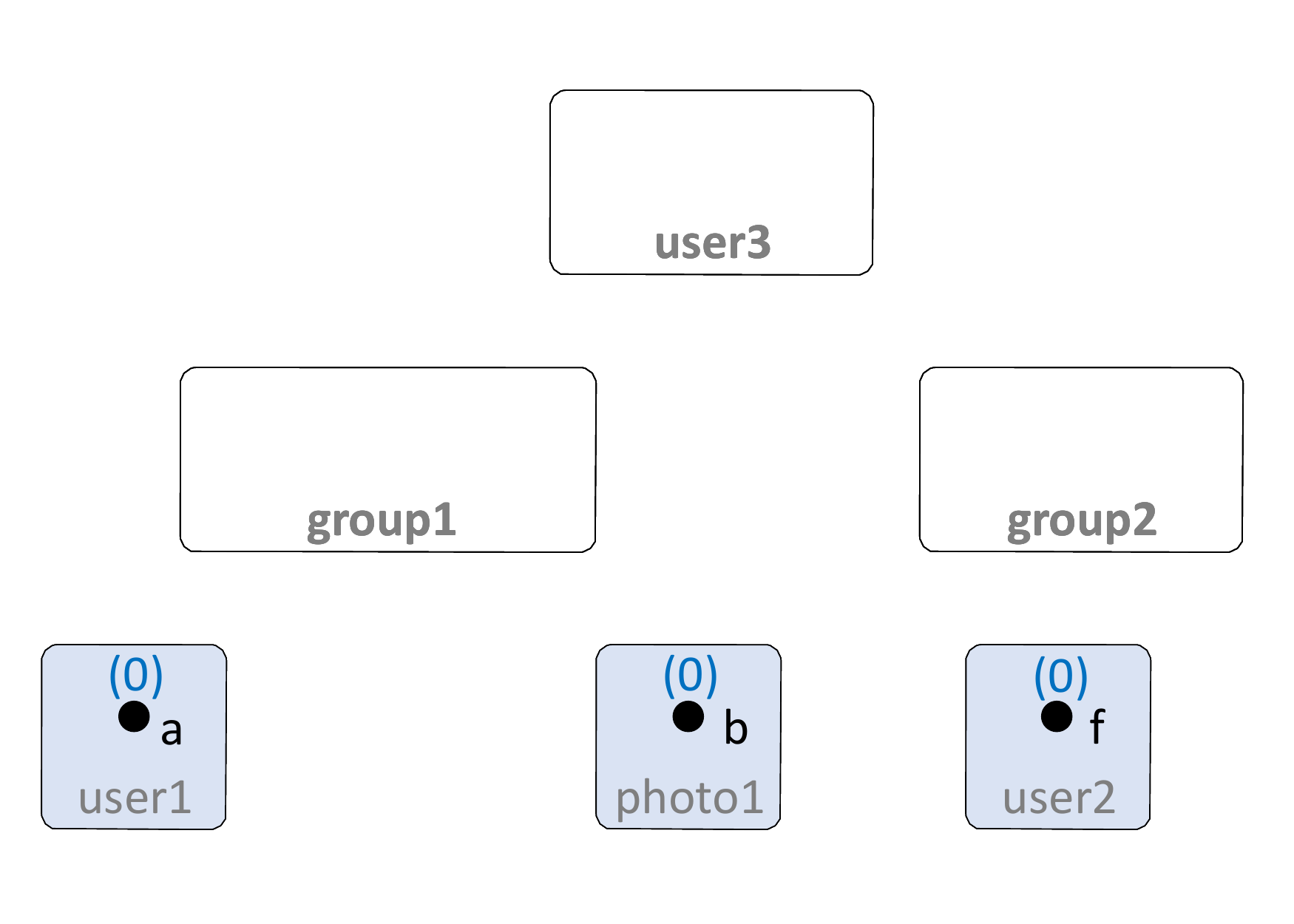}
	\caption{}\label{fig:bottomUp1}
\end{subfigure}
\hspace{3mm}
\begin{subfigure}[t]{.31\linewidth}
	\includegraphics[scale=0.31]{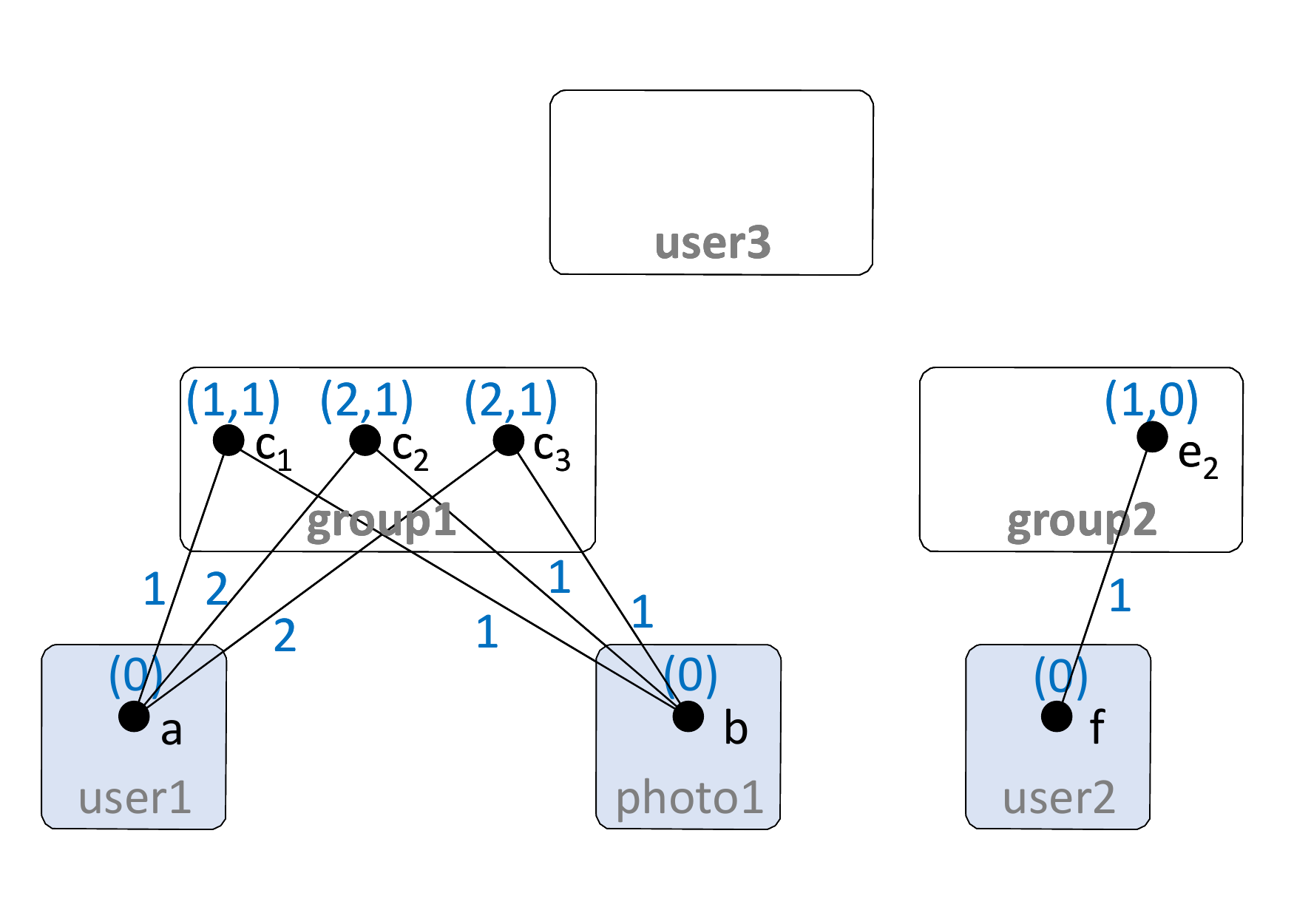}
	\caption{}\label{fig:bottomUp2}
\end{subfigure}
\hspace{3mm}
\begin{subfigure}[t]{.31\linewidth}
	\includegraphics[scale=0.31]{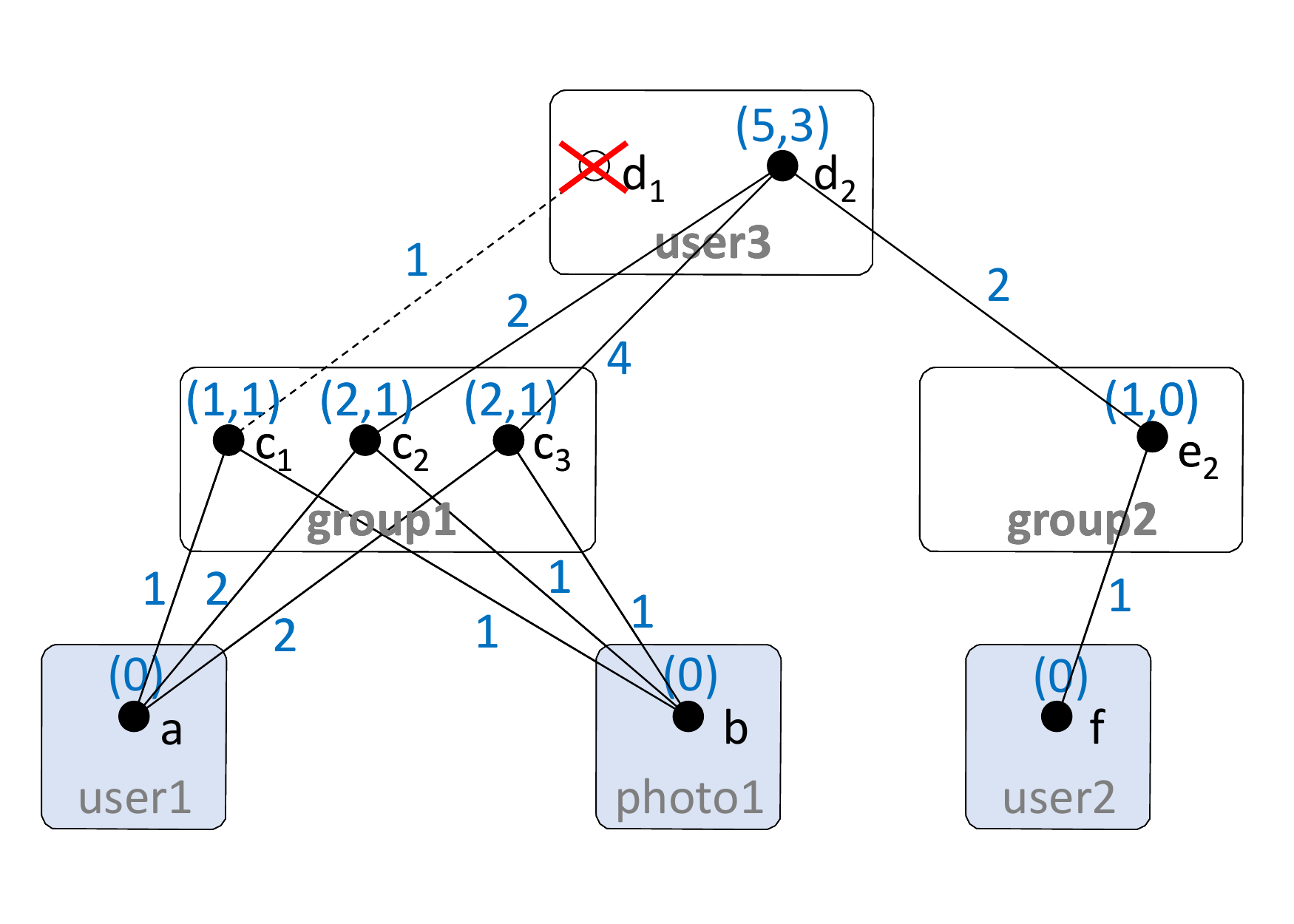}
	\caption{}\label{fig:bottomUp3}
\end{subfigure}

\caption{Minimal subtree weight computations: 
(a) after traversing all leaves,
(b) after traversing middle level,
(c) after finishing at the root.
Numbers above node candidates indicate minimum sub-tree weights stored in $\candnode$; numbers on edges indicate edge weights stored in $\candedge$.
}
\end{figure*}

The bottom-up phase traverses the query tree in any bottom-up order and
constructs a ``candidate graph'' consisting of two index structures:
(1) $\candnode(u)$ returns for query node $u$ a hash index that maps a node candidate $c$ to a list of minimum subtree weights, with one weight for each of $c$'s children.
(2) $\candedge(u, u')$ returns for each query edge between a node $u$ and its child $u'$ a hash index that maps a candidate node $c$ of $u$ to all adjacent candidates $c'$ of $u'$.

We illustrate 
Algorithm~\ref{alg:pruning} with Figures~\ref{fig:bottomUp1}, \ref{fig:bottomUp2}, and \ref{fig:bottomUp3}.
It first inserts candidate nodes for each query leaf node $u$ into the corresponding candidates $\candnode(u)$, 
setting their weights to zero (line~\ref{alg:line:terminal}). 
Note that leaves do not have children, hence the NIL value in the expression.
In Figure~\ref{fig:bottomUp1} there is a single candidate per leaf, but in practice it can be a larger subset of $\Vertices$ for each query leaf, 
depending on the node constraints. 
Then, for each query node $u$, the algorithm 
($i$) finds possible candidate nodes,
($ii$) prunes them, and  
($iii$) calculates the minimum subtree weights

In more detail: 
($i$) for each query edge leading to a child $(u,u')$, it first finds all candidate edges $(c, c')$, storing the map 
$\candedge: (u, u') \mapsto [c \mapsto c']$ (line~\ref{alg:line:createcandidates}).
($ii$) Then, the algorithm only keeps the list of candidates 
for each query node that are \emph{reachable from candidate instances in all leaves of the query node} (line~\ref{alg:line:intersection}):
In Figure~\ref{fig:bottomUp3}, the list of 
candidates for query node \textit{group1} is $\{c_1, c_2, c_3\}$. 
Notice how spurious candidates not reachable from the leaves, e.g., $e_1$ in \textit{group2}, are not even accessed (compare with Figure~\ref{fig:joinGraph}). 
Similarly, while $d_1$ in \textit{user3} is reachable from the left, it is not reachable from the right subtree and is thus automatically pruned as well.
($iii$)~Then, the algorithm finds for each reachable node, the min weight along each query edge $(u, u')$ starting at $c$ (line~\ref{alg:line:minsubtree}).
For example, in Figure~\ref{fig:bottomUp3},
the left weight $5$ for $c_2$ is computed 
as the minimum of weights 
for following $(d_2, c_2)$, 
which is 5 as the sum of the weight of edge $(d_2, c_2)$ (= 2)
plus the weight of $c_2$ (= 2+1),
or for following $(d_2, c_3)$, which is 7 as the sum of the weight of edge $(d_2, c_2)$ (= 4)
plus the weight of $c_3$ (= 2+1).
Notice we use here $\Priority(c)$ as short form for the sum of weights at a node $c$,
which we get from \candnode.
The two new created indices speed up finding adjacent edges in a subtree of the query pattern during top-down traversal.

\subsection{Top-Down Phase}

The second part of our algorithm performs top-down search, starting at the root node and proceeding downward to the leaves. This is essential for two reasons: First, the pre-computed subtree weights provide information to guide the search to the lightest patterns before exploring the heavier ones. Second, the top-down traversal implicitly prunes \emph{all} remaining spurious candidates for sub-graph homomorphism, as we will prove in Section~\ref{sec:analysis}. Again, pruning actually happens implicitly by not reaching those candidates. To see the latter, consider \textit{group1} candidate $c_1$ in Figure~\ref{fig:bottomUp3}. It is spurious, but could not be removed by the bottom-up sweep. However, it will never be accessed during top-down traversal, because $d_1$ was never recorded in \candnode by Algorithm~\ref{alg:pruning}.

\setlength{\textfloatsep}{0pt}
\begin{algorithm}[bt]
  \small
  \caption{\label{alg:prioritized-search}Prioritized Search (front-element optimization not shown)}
\begin{flushleft}  
\textbf{Input}: Tree pattern $\TreePattern$, \candnode, \candedge \\
\textbf{Output}: All matches of $\TreePattern$, one-by-one in increasing order of weight
\end{flushleft}
  \begin{algorithmic}[1]
	\State \algocomment{//Initialize \texttt{pq} with all candidates of the query root node}
	\State $\texttt{pq} \gets \textsc{PriorityQueue}()$
	\For{$c \in \candnode(\TreeRoot(\TreePattern))$}\label{line:alg:initpq}
    	\State $Z \gets$ partial tree $(c)$ consisting of just one node 
		\State $\texttt{pq}.\textsc{Insert}(\Priority(c), Z)$
    \EndFor
	\State \algocomment{//Expand \text{pq} until all results returned}
    \While{\texttt{pq}.\textsc{Size} > 0}
    	\State \label{line:update-cost} $(\texttt{oldkey}, Z) \gets \texttt{pq}.\textsc{Pop-Minimum}$
        \If{$Z$ is a complete match}
          \State return $Z$
		\Else
	        \State $(u, u') \gets \textsc{NextPreorder}(\TreePattern,Z)$ \label{line:alg:next}\Comment Edge to expand pattern
	        \For{$c \in \candnode(u)$}
	          \For{$c'$ returned by $\candedge(u,u').\textsc{Get}(c)$}
	            \State $Z' = Z.\textsc{Append}(c')$\label{line:expansion}
	            \State $\texttt{newkey} \gets 
					\texttt{oldkey} 
					- \candnode(u).\texttt{Get}(c, u')	
					+ 
					\phantom{a}\phantom{a}\phantom{a}\phantom{a}\phantom{a}\phantom{a}\phantom{a}
					\hspace{21mm}\phantom{a}
					w(c,c')
					+ \Priority(c')
					$
					\label{line:newPrio}			
	            \State \texttt{pq}.\textsc{Push}($\texttt{newkey}, Z')$
	          \EndFor
			\EndFor
        \EndIf
    \EndWhile
 \end{algorithmic}
\end{algorithm}

Algorithm~\ref{alg:prioritized-search} shows the pseudo-code for top-down guided search. 
Initially, all candidates $c$ in the query root $r$ are inserted into priority queue \texttt{pq} (line~\ref{line:alg:initpq}), 
with their priorities set to the sum of the candidate's weights.
In Figure~\ref{fig:bottomUp3}, there is a single candidate, $d_2$, of weight $5+3=8$. 
Then the algorithm repeatedly pops the top element from \texttt{pq} and expands the partial pattern using pre-order traversal. 
Function $\textsc{NextPreorder}$ returns the edge, as the pair of parent and child node, along which the partial pattern will be expanded next (line~\ref{line:alg:next}). 
The priority value of each expanded partial match is defined as the sum of the pattern's edge weights plus the sum of the weights of the unexplored subtrees. 
In the example, partial match $(d_2, c_2)$ is inserted into \texttt{pq} with priority 8 = 2 (edge weight) + (2+1) (weights of $c_2$) + 3 (weight of right subtree of $d_2$). 
Similarly, partial match $(d_2, c_3)$ is inserted with priority 4+(2+1)+3 = 10. 
Note that those values are computed incrementally during traversal (line~\ref{line:newPrio}). 
Consider expansion of $(d_2)$ to $(d_2, c_3)$. Priority of $d_2$ was 8, with weight 5 for the newly expanded subtree rooted at \textit{group1}. 
After retrieving $c_3$ from \candedge, priority of $(d_2, c_3)$ is computed as 
8 (\texttt{old})
- 5 (newly expanded subtree)
+ 4 (weight of edge $(d_2, c_3)$ )
+ 3 (priority of $c_3$) = 10 (line~\ref{line:newPrio} in Alg.~\ref{alg:prioritized-search}).
Then $(d_2, c_2)$ is popped next, and expanded to partial match $(d_2, c_2, a)$ with priority 
8 = 8 - 2 + 2 + (0+0).
This pattern is then expanded next to $(d_2, c_2, a, b)$, $(d_2, c_2, a, b, e_2)$, and finally $(d_2, c_2, a, b, e_2, f)$---all with the same priority of 8. The latter is output as the minimal-weight solution. Only then will partial match $(d_2, c_3)$ with the higher priority value 10 be expanded analogously. Each expansion operation requires a pop operation from priority queue, visiting potential edges once.

\section{Algorithm Analysis}
\label{sec:analysis}

All results in this section are for the relaxed version of the problem, based on sub-graph homomorphism instead of isomorphism. We discuss in Section~\ref{sec:repetitions} how to extend them to the isomorphism case. Proofs were omitted due to space constraints, but can be found in the extended version~\cite{yang2018}.

\subsection{Minimality of Candidate Graph}

We show that during top-down search (Alg.~\ref{alg:prioritized-search}), no spurious candidate node will ever be accessed. A candidate node $c$ for a query node $q$ is ``\emph{spurious}'' if there does not exist any homomorphic result pattern where $c$ is matched to $q$. Ensuring that no spurious nodes are accessed is crucial for proving strong upper bounds on the algorithm cost.

\begin{theorem}\label{thm:spuriousNode}
If node candidate $c \in \candnode(q)$ for query node $q \in V_Q$ is accessed by Alg.~\ref{alg:prioritized-search}, then there exists a homomorphic result pattern where $\BijectionFunct(q) = c$.
\end{theorem}

\subsection{Each Pop, One Result---In Order}
\label{sec:pq_pops}

Next, we show a powerful result that is crucial in establishing important algorithm properties: 
During the top-down guided search, for each query result there is \emph{at most} one push and \emph{at most} one pop operation on priority queue \texttt{pq}.
For this, we need the following lemmas.

\begin{lemma}\label{lem:priorityMonotonicity}
The priority value of a partial pattern $P$ is always less than or equal to the priority of all its successors.
\end{lemma}

\begin{lemma}\label{lem:sameWeightPQ}
Assume that Alg.~\ref{alg:prioritized-search} popped partial pattern $P=(c_1, c_2,\ldots, c_j)$, $j < |V_Q|$, of priority $w$ from \texttt{pq}. Then there exists a direct successor $(c_1, c_2,\ldots, c_j, c_{j+1})$ that has the same priority $w$.
\end{lemma}

Lemmas~\ref{lem:priorityMonotonicity} and \ref{lem:sameWeightPQ} immediately imply:

\begin{corollary}\label{cor:frontOptOk}
If the last pop operation on \texttt{pq} returned an incomplete pattern $P$, then one of the direct successors of $P$ will have priority equal to the minimum priority over all elements in \texttt{pq}.
\end{corollary}

\begin{example}\label{ex:PETpriorityValues}
Consider the changes of \texttt{pq} for the example in Figure~\ref{fig:bottomUp3}. Initially it contains $[(d_2):8]$, the sole root node candidate with priority 5+3=8. This element is popped and expanded along edges $(d_2, c_2)$ and $(d_2, c_3)$. The priority of the former is 2 (weight of edge $(d_2, c_2)$) plus (2+1) (subtree weights of $c_2$) plus 3 (right subtree weight of $d_2$) = 8. It is identical to the initial priority of $d_2$, because edge $(d_2, c_2)$ is the one that determined the minimum left subtree weight of 5 in $d_2$. For $(d_2, c_3)$, priority is 10 due to the higher weight of edge $(d_2, c_3)$. After these two patterns are pushed, \texttt{pq} contains $[(d_2, c_2):8, (d_2, c_3):10]$. The next pop delivers $(d_2, c_2):8$, which is expanded to $(d_2, c_2, a):8$, followed by repeated pop and push operations on this pattern, every time obtaining the same priority of 8, until the top result $(d_2, c_2, a, b, e_2, f)$ of weight 8 is completed. Only then will expansion of $(d_2, c_3):10$ commence.
\end{example}

\introparagraph{Front-element optimization}
Based on Corollary~\ref{cor:frontOptOk}, we next introduce an important optimization to Alg.~\ref{alg:prioritized-search}. Since the corollary guarantees that one of the direct successors of the partial pattern popped before will have a minimal priority value, we avoid the push-pop cycle for it and keep expanding it directly, only pushing the other direct successors. More precisely, assume the algorithm just popped partial match $P=(c_1, c_2,\ldots, c_i)$ of priority $w$ from \texttt{pq}. While expanding this pattern by one more node, it keeps in memory the first direct successor $P'=(c_1, c_2,\ldots, c_i, c'_{i+1})$ encountered that also has priority value $w$, pushing all other direct successors to \texttt{pq}. This way the algorithm still works on a min-priority element, but avoids the push-pop cycle for it. This seemingly minor optimization has strong implications as formalized in the following theorems.

\begin{theorem}\label{thm:kPop}
Using front-element optimization, for any $k$, the $k$-th pop operation from \texttt{pq} produces the $k$-th lightest homomorphic result pattern, possibly requiring additional push operations, but no more pop operations until this result pattern is returned.
\end{theorem}

\begin{corollary}\label{cor:numPushes}
No matter how many results are retrieved, Alg.~\ref{alg:prioritized-search} never performs more than $r_H$ push operations on \texttt{pq} \emph{in total}. Here $r_H$ denotes the number of homomorphic subtrees in $\Graph$.
\end{corollary}

This follows directly from Theorem~\ref{thm:kPop} and the following observation. Assume the algorithm continues to run until all query results are found. At that point it has removed all partial matches from \texttt{pq} and the queue is empty. Theorem~\ref{thm:kPop} implies that retrieving all results requires exactly $r_H$ pop operations. If the total number of push operations exceeded this, then the queue would not be empty. (And obviously, any execution of Alg.~\ref{alg:prioritized-search} that stops before returning all results will only have performed a subset of the push operations executed by the time all results are returned.)

\subsection{Algorithm Cost}
\label{sec:algCost}

To avoid notational clutter, we treat the size of the query pattern as a small constant and omit it from most formulas. (Note that pattern size is equal to the number of edges in $\EdgesPattern$, e.g., 5 in the photo-sharing network example.) It is straightforward to extend the formulas by including $|\EdgesPattern|$ as a variable.

\introparagraph{Algorithm~\ref{alg:pruning}}
Theoretical worst case cost is $\mathrm{O}(|\Edges|)$\hide{$\mathrm{O}(|\Edges| \cdot |\EdgesPattern|)$}, i.e., linear in graph size: for each of the query pattern edges, in the worst case all graph edges are accessed. The time for constructing $\candedge$ and $\candnode$ adds a constant overhead per edge processed. In practice, only a small fraction of $\Edges$ will be accessed because of the label constraints. In particular, by using $\graphedge$ in line~\ref{alg:line:createcandidates} in Alg.~\ref{alg:pruning}, all neighbors of matching types (labels) are accessed in time linear in the number of these neighbors. Space cost is upper bounded by the combined size of $\candnode$ and $\candedge$, i.e., cannot exceed $|\EdgesPattern|$ times input graph size.

\introparagraph{Algorithm~\ref{alg:prioritized-search}}
The results from Section~\ref{sec:pq_pops} lead to strong guarantees. Space complexity of Alg.~\ref{alg:prioritized-search} is equal to the maximum size of the priority queue.\hide{Since partial matches grow during the search process, we measure queue size not just in terms of the number of elements, but also consider their size---determined by the number of matched nodes.} Corollary~\ref{cor:numPushes} immediately implies:

\begin{theorem}\label{thm:spaceAlg2}
Space cost of Alg.~\ref{alg:prioritized-search} is upper bounded by $r_H$\hide{$r_H \cdot |\EdgesPattern|$}, the total result size for sub-graph homomorphism.
\end{theorem}

From a user's point of view, the time it takes to produce the next lower-ranked result is crucial:

\begin{theorem}\label{thm:interarrival}
The initial latency for Alg.~\ref{alg:prioritized-search} to return the top-ranked homomorphic match, and also the time between returning any two consecutive homomorphic matches, is $\mathrm{O}(\mathrm{outDegree} + \log r_H)$. 
Here $\mathrm{outDegree} \le r_H$ is greater of (1) the number of candidates in the root node and (2) the maximum cardinality of the set of adjacent node candidates $c'$ in $\candedge: (u, u') \mapsto [c \mapsto \{(c', \WeightFunct(c, c'))\}]$ for any query graph edge $(u, u')$ and candidate $c$.
\end{theorem}

These strong results show that {\ouralgorithmname} can effectively exploit selective label constraints. For instance, if there are a thousand homomorphic subgraphs in $\Graph$, then Theorems~\ref{thm:spaceAlg2} and \ref{thm:interarrival} guarantee that Alg.~\ref{alg:prioritized-search} will never store more than a thousand partial matches in memory and will perform at most a thousand (inexpensive) computation steps to deliver the next result to the user---no matter how big or connected the given graph!

We show next that the anytime property of {\ouralgorithmname}, i.e., that it can deliver the top-ranked results quickly and then the next ones on request, incurs \emph{no performance penalty} for producing \emph{all} homomorphic matches:

\begin{theorem}\label{thm:timeAlg2}
The lower bound for producing \emph{all} homomorphic result patterns is $\Omega(r_H)$; sorting them costs $\mathrm{O}(r_H \log r_H)$. Alg.~\ref{alg:prioritized-search} has matching total time complexity $\mathrm{O}(r_H \log r_H)$\hide{$\mathrm{O}(|E_C| + r_H \log r_H)$}.
\end{theorem}

\section{Homomorphism to Isomorphism}
\label{sec:repetitions}

{\ouralgorithmname} as introduced in Section~\ref{sec:algorithm} returns homomorphic matches. To obtain the desired isomorphic matches, function $\BijectionFunct$ mapping query nodes to tree-pattern nodes has to be bijective. To guarantee this, one simply has to filter out all results where different query nodes are mapped to the same graph node. Instead of filtering on the final result, {\ouralgorithmname} can perform early pruning by checking in line~\ref{line:expansion} in Alg.~\ref{alg:prioritized-search} if newly added node $c'$ already appears in partial match $Z$---discarding $Z'$ if it does. This modification has the following implications for the cost analysis results in Section~\ref{sec:algCost}.

Since some of the items previously pushed to priority queue \texttt{pq} will now be discarded early, space consumption as well as computation cost of {\ouralgorithmname} are lower than for finding all subgraph homomorphism results. However, worst-case complexity as established by Theorems~\ref{thm:spaceAlg2} and \ref{thm:timeAlg2} remains the same. And the guarantees for the time between results (Theorem~\ref{thm:interarrival}) is weaker: In the worst case, e.g., when only the very first and the very last of the homomorphic matches represent isomorphic results, then time between consecutive results grows from $\mathrm{O}(\mathrm{outDegree} + \log r_H)$ to $\mathrm{O}(r_H \log r_H)$. Fortunately, as our experiments indicate, heterogeneity indeed results in a small gap between homomorphism and isomorphism, i.e., real-world performance is closer to $\mathrm{O}(\mathrm{outDegree} + \log r_H)$.

\section{Experiments}
\label{sec:exp}

Our experiments evaluate running time, memory consumption, and size of the search space explored by $\ouralgorithmname$ on several query templates and four data sets against two baseline algorithms. In order to allow reproducible results, our code can be downloaded from our project page~\cite{Code2018}.

\introparagraph{Baseline algorithms}
We chose two baselines that allow us to evaluate the relative contribution of our two key steps 
(pruning the search space, and guided search) 
to the performance of $\ouralgorithmname$.

\emph{\baselineone} first calculates all results from the candidate graph and only then ranks them. 
Since it uses our pruned search space, but does not include any prioritizing of query results, it serves as a baseline to evaluate the contribution of our guided search phase.

\emph{\baselinetwo} is intended to evaluate the effect of our aggressive tree-based pruning strategy. It extends the state-of-art top-$k$ algorithm for \emph{path} queries 
on HINs~\cite{liang2016links}. First, it identifies the longest terminal-to-terminal path $R$ in the tree pattern $\TreePattern$ (called the ``\emph{backbone}'' of $\TreePattern$). 
It then incrementally retrieves the lightest backbone instances in $\Graph$ one-by-one. Note that such an instance is a partial match with smaller unmapped subtrees ``hanging off'' the backbone path. Thus, we can execute {\baselineone} on each such subpattern, yielding a divide and conquer algorithm. Since each subpattern is independent, we extract the lightest instance of each subpattern, and merge these solutions with the backbone instance. This involves checking for repeat node occurrences to enforce isomorphism. Then the next heavier subtree matches are explored etc. Given a pre-defined value of $k$, the algorithm can prune the set of remaining backbone instances every time a new full match is found. \emph{In all our experiments, we supply the ultimate value of $k$ to {\baselinetwo}, to explore the best possible performance this algorithm might achieve (if it was able to guess the correct value of $k$ from the start).}

\introparagraph{Datasets}
We use four well-known heterogeneous datasets: 
Flickr \cite{Flicker},
DBLP \cite{liang2016links},
Enron \cite{liang2016links},
and
Yelp \cite{Yelp}. 
Figure~\ref{fig:datasets} gives an overview of their properties.
Notice that Enron, DBLP and Flickr have denser graphs than Yelp.
In order to create weights for individual edges, we used the age of the edge with an exponential decay based on the difference between edge creation time and query time.

\begin{figure}
\begin{center}
    \begin{tabular}{| l | r | r | c |}
    \hline
    Dataset & \multicolumn{1}{c|}{$|\Vertices|$} &\multicolumn{1}{c|}{$|\Edges|$} &\multicolumn{1}{c|}{$|\LabelSet|$}  \\ \hline\hline
    Flickr &2,007,369 &18,147,504 &3  \\ \hline
    DBLP &2,241,258 &14,747,328 &4  \\ \hline
    Enron &46,463 &613,838 &4 \\ \hline
    Yelp &4,301,900 &7,059,472 &6  \\ \hline
    \end{tabular}
\end{center}
\caption{Dataset statistics}
\label{fig:datasets}
\end{figure}

\begin{figure}[t]
    \includegraphics[width=0.45\textwidth]{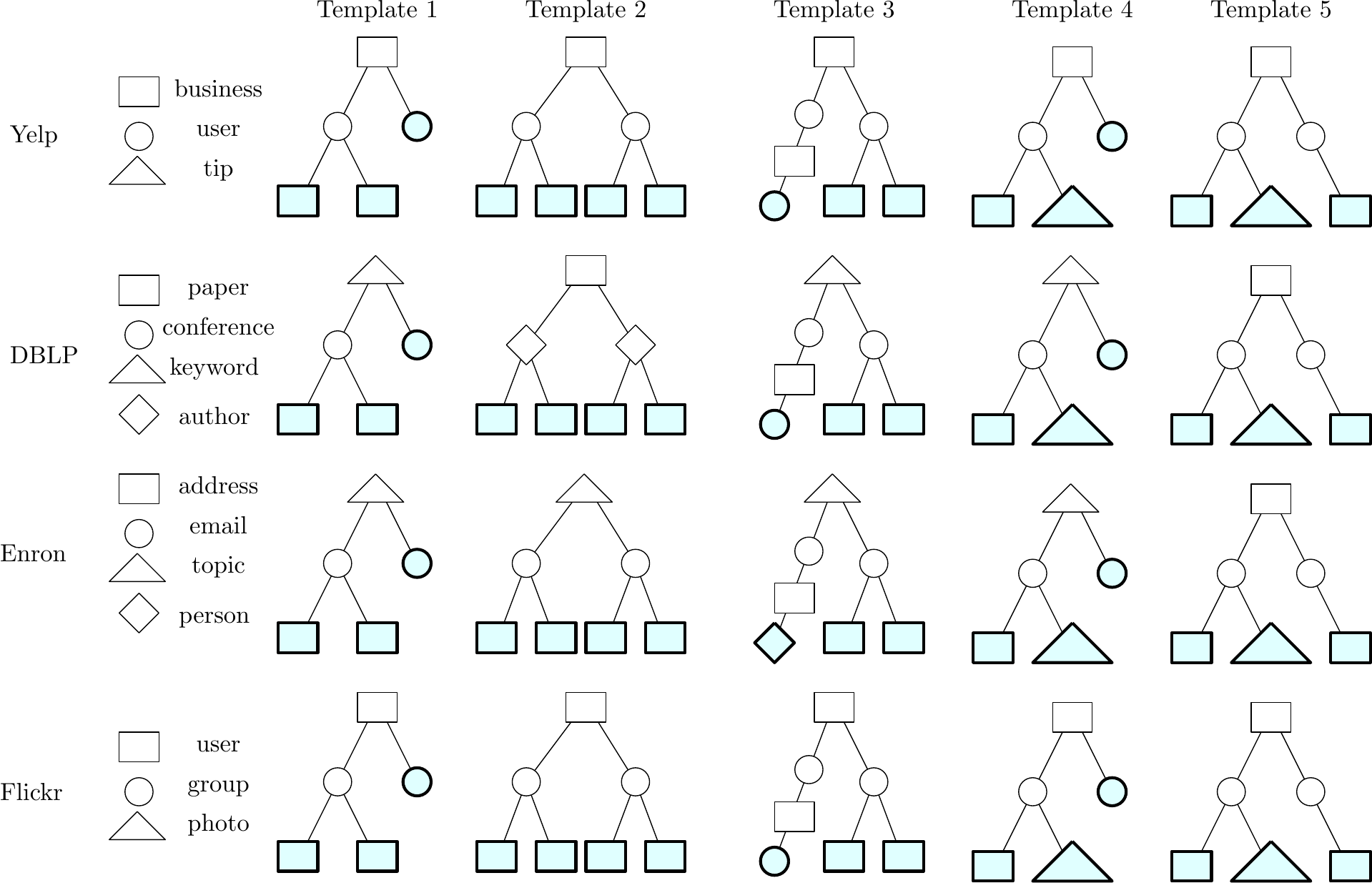}
    \caption{Templates used to generate queries for each dataset. Shapes indicate node types; the cyan-colored nodes are terminals.}
    \label{fig:templates}
\end{figure}

\introparagraph{Query templates}
We created five different pattern \emph{templates} listed in Fig.~\ref{fig:templates}.
For each pattern, we chose 300 different assignments of actual nodes to the query leaves, which gave us 300 queries per template, for a total of 6,000 query instances across all datasets.

\introparagraph{Performance metrics}
We vary $k$ between 1 and total number of results, and compare the algorithms on three performance metrics. 
1) \emph{Running time}: We measure the time between having loaded the dataset into memory and returning the k-th result, reporting the average for each template.
2) \emph{Memory consumption}: We report the \emph{maximum} number of nodes (i.e., partial matches weighted by the number of matched nodes they contain) stored at any time during the execution of each algorithm.
3) \emph{Size of search space}: We report the \emph{total} number of partial matches (weighted by the number of matched nodes they contain) that are stored during execution.

\introparagraph{Experimental setup}
Experiments were run on an Intel Xeon CPU E5-2440 1.90GHz with 200GB of memory running Linux. We compiled the source code with \texttt{g++} 4.8 (optimization flag \texttt{O2}).

\subsection{\label{sec:results}\textbf{Discussion and Highlights}}

\begin{figure}[t]
\centering
\begin{subfigure}[t]{0.4\textwidth}
  \includegraphics[width=\textwidth]{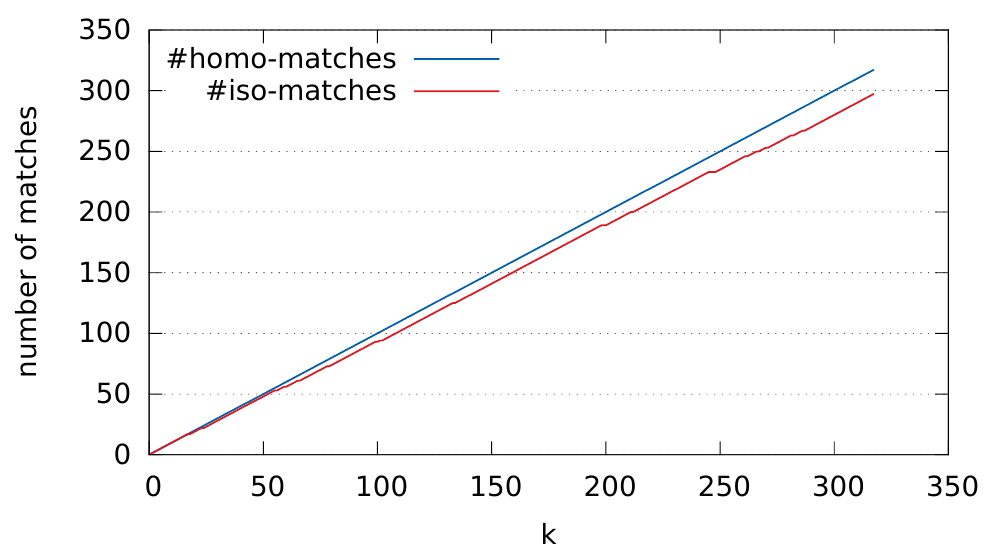}  
\end{subfigure}
\caption{Number of isomorphic and homomorphic matches for an instance of Template 1 on the Enron dataset.}\label{fig:homo_iso}
\end{figure}

To justify our ``solving isomorphism through homomorphism'' approach, we also ran {\ouralgorithmname} with node-repetition check turned off. This produces all homomorphic matches, from which we can then determine offline how many were also results for the isomorphism case. We plot both numbers as we increase $k$ in Figure~\ref{fig:homo_iso}. It shows a representative result, obtained from the Enron dataset using a query from Template~1. The small gap between the lines confirms that the vast majority of homomorphic matches are also isomorphic result patterns.

\begin{figure}[t]
\centering
\begin{subfigure}[t]{0.4\textwidth}
  \includegraphics[width=\textwidth]{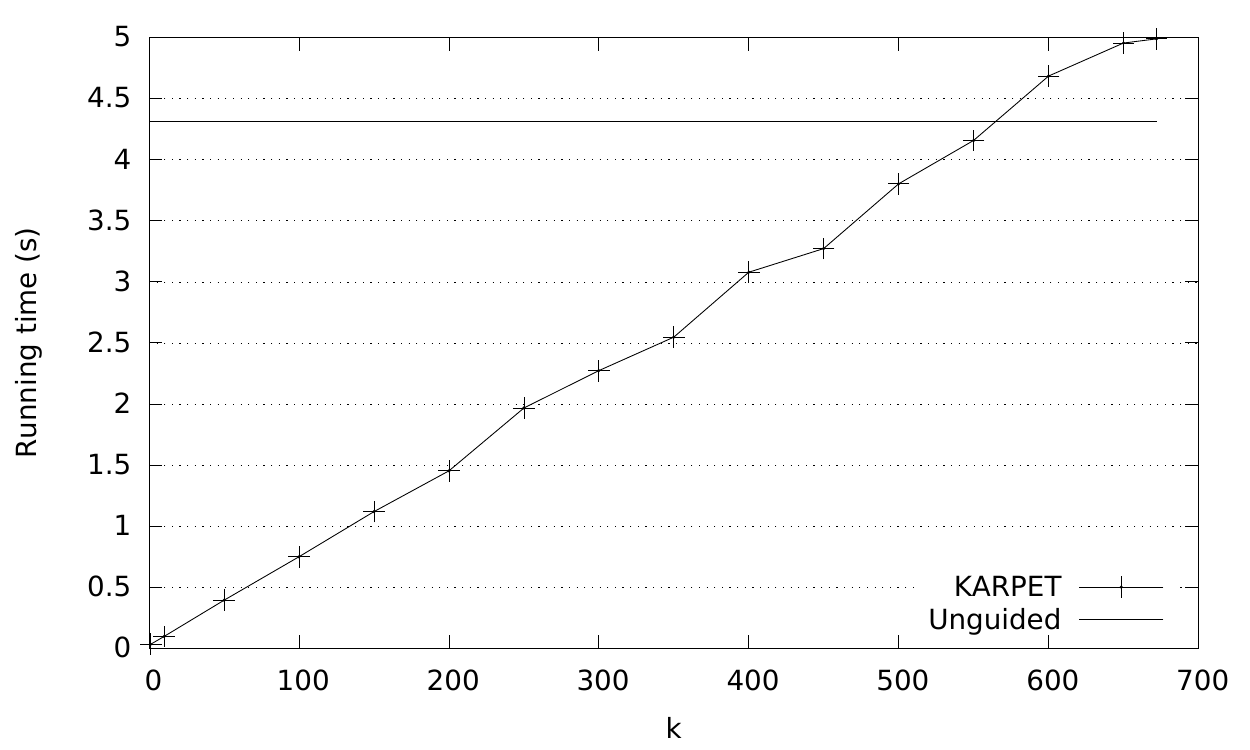}
\end{subfigure}
\caption{\label{fig:topk}Running time for increasing k until all results are returned, for {\ouralgorithmname} and {\baselineone}.}

\end{figure}

Figure~\ref{fig:topk} shows a representative result comparing {\ouralgorithmname} to {\baselineone}, which bulk-computes the entire output and hence is not affected by the choice of $k$. It is clearly visible how our any-$k$ algorithm continuously returns the top-ranked results in order, with very low latency between consecutive outputs. By the time {\baselineone} finally returns the first match, {\ouralgorithmname} has already delivered more than 85\% of all matches. In the end, it took {\ouralgorithmname} 4.98s to output \emph{all} matches. For comparison, bulk-computation by {\baselineone} took 4.31s, indicating a small overhead of 0.67s for supporting the anytime property.

\begin{table}[bt]
\npdecimalsign{.}

\footnotesize
\begin{tabular}{|l|l|r|r|r|r|r|}
\hline
{Data} &{Algorithm} & { \: T1  \:} & {  \:T2 \: } & {  \:T3 \: } & {  \:T4 \: } & {  \:T5 \: } \\
\hline\hline
\multirow{3}{*}{Flickr} & {KARPET} &43.86 \cellcolor[gray]{.8}&58.17 \cellcolor[gray]{.8}&99.28 \cellcolor[gray]{.8}&20.12 \cellcolor[gray]{.8}&81.92 \cellcolor[gray]{.8}\\ 

{} &{Unguided} &2200 &2085 &5951 &2021 &5518 \\ 

{} & {Backbone} &2946 &9678 &3033 &4114 &9963 \\\hline\hline 

\multirow{3}{*}{DBLP} &{KARPET} &9.51 \cellcolor[gray]{.8}&8.87 \cellcolor[gray]{.8}&104.33\cellcolor[gray]{.8} &8.67\cellcolor[gray]{.8} &8.15 \cellcolor[gray]{.8}\\ &{Unguided} &1090 &1936 &1995 &814 &1110 \\ 

&{Backbone} &839 &502 &337 &1109 &1059 \\ 

\hline\hline 

\multirow{3}{*}{Enron} &{KARPET} &10 \cellcolor[gray]{.8}&57\cellcolor[gray]{.8} &29\cellcolor[gray]{.8} &57 \cellcolor[gray]{.8}&104\cellcolor[gray]{.8} \\ 

{} & {Unguided} &202 &1036 &1013 &669 &891 \\ 

{} & {Backbone} &4914 &18925 &5124 &7056 &11075 \\ 

\hline\hline 
\nprounddigits{3}
\multirow{3}{*}{Yelp} & {KARPET} &0.11\cellcolor[gray]{.8} &0.13 \cellcolor[gray]{.8}&0.78 \cellcolor[gray]{.8}&0.20\cellcolor[gray]{.8} &0.15\cellcolor[gray]{.8} \\ 

{} & {Unguided} &1.68 &1.79 &2.27 &1.40 &2.06 \\ 

{} & {Backbone} &3.84 &3.53 &1.02 &3.82 &3.02 \\ 
\hline
\end{tabular}
\caption{\label{tab:running-time}Average running time (milliseconds) for each dataset, algorithm, and template. Cells with the least running time in the three algorithms are marked gray.}
\end{table}

\begin{table}[bt]
\npdecimalsign{.}
\nprounddigits{3}
\footnotesize

\begin{tabular}{|l|l|r|r|r|r|r|}
\hline
{Data} &{Algorithm} & { \: T1  \:} & {  \:T2 \: } & {  \:T3 \: } & {  \:T4 \: } & {  \:T5 \: } \\
\hline\hline

 \multirow{3}{*}{Flickr} &{KARPET}  &278 \cellcolor[gray]{.8}&286   &1212  &458  \cellcolor[gray]{.8}&1957 \\
 {} &{Unguided}  &1212 &294  &17560 &833 &61603\\ 
 {} &{Backbone}  &460  &139\cellcolor[gray]{.8}  &76 \cellcolor[gray]{.8}&939&1057 \cellcolor[gray]{.8}\\
 \hline\hline
 \multirow{3}{*}{DBLP} &{KARPET} &162 &146 &387 &251 &253 \\
 &{Unguided} &162 &146 &409 &251 &253 \\
  &{Backbone} &29\cellcolor[gray]{.8} &35\cellcolor[gray]{.8} &26 \cellcolor[gray]{.8}&72\cellcolor[gray]{.8} &103\cellcolor[gray]{.8} \\
 \hline\hline
 \multirow{3}{*}{Enron} &{KARPET} &459 &676 &546 &727 &713 \\
 {} &{Unguided} &477 &697 &569 &745 &728 \\
 {} &{Backbone} &188 \cellcolor[gray]{.8}&109 \cellcolor[gray]{.8}&11 \cellcolor[gray]{.8}&168\cellcolor[gray]{.8} &87 \cellcolor[gray]{.8}\\
 \hline\hline
 \multirow{3}{*}{Yelp} &{KARPET} &1.02 &1.02 &1.36 &1.01 &1.02 \\
 {} &{Unguided} &1.02 &1.02 &1.36 &1.01 &1.02 \\
 {} &{Backbone} &1.03 &1.02 &1.00\cellcolor[gray]{.8}&1.01 &1.02 \\
\hline
\end{tabular}
\caption{\label{tab:memory} Average memory measured by the max number of nodes stored, for each dataset, algorithm, and template. Cells using the least memory are marked gray.
}
\end{table}

\begin{table}[bt]
\npdecimalsign{.}
\nprounddigits{3}
\footnotesize
\begin{tabular}{|l|l|r|r|r|r|r|}
\hline
{Data} &{Algorithm} & { \: T1  \:} & {  \:T2 \: } & {  \:T3 \: } & {  \:T4 \: } & {  \:T5 \: } \\
\hline \hline
 \multirow{3}{*}{Flickr} & {KARPET}&1.16-e6  \cellcolor[gray]{.8}&1.18-e6 \cellcolor[gray]{.8}&9.29-e5  \cellcolor[gray]{.8}&1.12-e4 \cellcolor[gray]{.8}&1.56-e6 \cellcolor[gray]{.8}\\
{} &{Unguided} &5.41-e6 &1.55-e6 &1.96-e7 &1.52-e5 &2.20-e7\\ 
{} & {Backbone} &5.37-e6 &1.99-e6  &1.16-e6  &1.02-e7 &8.04-e6 \\

 \hline\hline
 \multirow{3}{*}{DBLP} &{KARPET} &4.84-e4 \cellcolor[gray]{.8}&3.98-e4  \cellcolor[gray]{.8}&5.48-e4  \cellcolor[gray]{.8}&9.35-e4  \cellcolor[gray]{.8}&4.55-e4  \cellcolor[gray]{.8}\\
 &{Unguided} &7.25-e5 &3.35-e5 &1.58-e6 &2.16-e5 &8.91-e5 \\
  &{Backbone} &4.64-e5 &1.68-e5 &2.91-e5 &9.38-e5 &9.57-e5 \\
 \hline\hline
 \multirow{3}{*}{Enron} &{KARPET} &4.22-e4  \cellcolor[gray]{.8}&1.13-e5 \cellcolor[gray]{.8}&7.39-e4  \cellcolor[gray]{.8}&1.29-e4 \cellcolor[gray]{.8}&3.52-e4  \cellcolor[gray]{.8}\\
 {} & {Unguided} &8.41-e5 &9.50-e5 &9.11-e5 &8.93-e5 &8.60-e5 \\
 {} & {Backbone} &1.31-e6 &8.86-e5 &1.39-e6 &3.07-e5 &5.74-e5 \\
 \hline\hline
 \multirow{3}{*}{Yelp} & {KARPET} &58&99 \cellcolor[gray]{.8}&191 &51 &104 \\
 {} & {Unguided} &58 &99 &233 &51 &104 \\
 {} & {Backbone} &45 \cellcolor[gray]{.8}&100 &17 \cellcolor[gray]{.8}&48  \cellcolor[gray]{.8}&62  \cellcolor[gray]{.8}\\
\hline
\end{tabular}
\caption{\label{tab:space} Average search space measured by weighted total number of partial matches, for each dataset, algorithm, and template. Cells exploring the least search space are marked gray.
}
\end{table}

For each template (fixing $k = 5$) we show the average running time, memory usage, and search space of our algorithm, along with the two baselines, on all four datasets in Tables~\ref{tab:running-time}, \ref{tab:memory}, and \ref{tab:space}. In terms of average running time, {\ouralgorithmname} outperforms the baselines in all cases. On DBLP, we can see 100x faster running time compared to {\baselineone}, and 10x speedup compared to {\baselinetwo}. We make the following detailed observations about these results.

\subsubsection{\textbf{Varying the Dataset}}

We observed that {\ouralgorithmname}'s margin of improvement over the baselines is generally greater on denser graphs. Higher density leads to more matches, which {\ouralgorithmname} can handle best, because it prioritizes the search based on subtree weight more effectively than the two baselines.
The \texttt{Yelp} graph is extremely sparse, causing many queries to have only one or two matches. For those queries, all three algorithms behave nearly identically, e.g., running time for a typical single-match query on \texttt{Yelp} was $0.15$ msec for {\ouralgorithmname}, $0.2$ msec for {\baselineone}, and $0.16$ msec for {\baselinetwo}.
However, even for this sparse graph, there are several queries for each pattern that have a larger number of candidate instances. 
These result in a significantly slower average running time for both baselines, while {\ouralgorithmname} averages to less than 1 msec.

{\baselinetwo} does relatively well compared to {\baselineone} on the larger dataset DBLP, and is the same order of magnitude for Flickr, but fails to perform well on Enron. For dense graphs, if the branch-and-bound does not terminate quickly, the overhead required by the divide-and-conquer merging steps can be very large because the same subtree may be visited multiple times.

\subsubsection{\textbf{Memory}}

The backbone-based algorithm often uses the least amount of memory, because in each iteration, it only holds the backbone matches and the single instance of the backbone it grows. The memory bottleneck for {\baselineone} is the amount of storage needed to hold all matching trees when they are sorted by weight.

\subsubsection{\textbf{Relative Strengths and Weaknesses}}

For ``easy'' queries, which only have a few matching instances, all three approaches show similar running time. On the other hand, when a query has to select the top-$k$ from a larger result set, e.g., dozens or 100s of results, {\ouralgorithmname} has a significant advantage from efficiently pruning the search space at an earlier stage. For such ``hard'' queries, {\baselineone} enumerates all matching instances before sorting them. {\baselinetwo} only partially exploits pruning opportunities for the backbone. The non-trivial extensions proposed for {\ouralgorithmname} are required to fully benefit from the constraints encoded by the entire tree structure.

{\baselinetwo} can exhibit faster running times than {\baselineone} in some cases. It achieves this speed-up due to its branch-and-bound nature, filtering out instances that exceed the threshold established by matches for the lightest backbones explored early on. (Note that this type of pruning takes advantage of advance knowledge of the final value of $k$. In practice, the algorithm would not know $k$ and hence could not apply any such pruning.) Either way, in most cases, this advantage is outweighed by the fact that {\baselinetwo} introduces an overhead for merging subtrees and repeatedly visiting some of the subtrees.

\section{Related Work}
\label{sec:related}

Our proposed notion of an any-$k$ algorithm is novel and extends the functionality of the previously-studied class of top-$k$ algorithms.
The problem of fast graph pattern search has been studied in different research communities, such as algorithms,
graph databases, and data mining.
While traditional data mining and work in theory focuses on the structure of the graph, meta-path based approaches~\cite{sun2011pathsim} also leverage the type information. 

\introparagraph{Subgraph isomorphism}
Subgraph isomorphism is an NP-hard problem~\cite{lubiw1981some}, and state-of-art algorithms 
are not practical for large graphs.
Lee \etal \cite{lee2012depth} empirically compare the performance of several state-of-art subgraph isomorphism algorithms, including the Generic Subgraph Isomorphism Algorithm~\cite{cordella2004sub}, Ullmann algorithm~\cite{ullmann1976algorithm}, VF2~\cite{cordella2001improved}, QuickSI~\cite{shang2008taming}, GADDI~\cite{zhang2009gaddi}, and GraphQL~\cite{he2008graphs}. They test on real-world datasets AIDS, NASA, Yeast and Human, covering a spectrum of relatively small graphs ($|\Vertices|<1000$).
Modern social networks easily exceed that size by orders of magnitude, and the exact sub-graph isomorphism problem remains intractable for larger networks when label constraints and top-$k$ are not fully exploited. Hence, to the best of our knowledge, none of these existing \emph{precise} pattern matching algorithms could be used for our target application.

\introparagraph{K-shortest simple paths}
The $k$-shortest paths problem is a natural and long-studied generalization of the shortest path problem, in which not one but several paths in increasing order of length are sought. The additional constraint of ``simple'' paths requires that the paths be without loops. This problem has been researched for directed graphs~\cite{doi:10.1287/mnsc.17.11.712}, undirected graphs~\cite{katoh}, approximation algorithms~\cite{Roditty07}, and algorithm engineering~\cite{Hershberger:2007:FKS:1290672.1290682,Feng14}. 
However, this body of literature was developed for graphs without labels. In contrast, {\ouralgorithmname} efficiently finds $k$-lightest instances matching a given query pattern by leveraging the heterogeneity constraints on the node and edge types to speed up the computation. Furthermore, in our scenario, $k$ is not known upfront.

\introparagraph{Querying graph data}
In the database community, querying and managing of large-scale graph data has been studied~\cite{Aggarwal2010:graphSurvey}, \eg in the context of GIS, XML databases, bioinformatics, social networks, and ontologies. The main focus there has been on identifying connection patterns between the graph vertices~\cite{faloutsos2004fast,tong2006center,wei2010efficient,kasneci2009ming,ramakrishnan2005discovering,cheng2009efficient}. In contrast, {\ouralgorithmname} finds matches for a given query pattern.

\introparagraph{HINs and path patterns}
Heterogeneous Information Networks (HINs)~\cite{han2010mining} are an abstraction to represent graphs whose nodes are affiliated with different types. To derive complex relations from such information networks, ``meta-paths'' defined as node-typed paths on a heterogeneous network, are a representation of connections between nodes, by specifying the types along a path in the network~\cite{sun2011pathsim}. Thus, while the focus in that line of research has been on learning good meta-path patterns for various applications,  {\ouralgorithmname} efficiently finds matches for a given query pattern.

Liang \etal \cite{liang2016links} derived a top-k algorithm for ranking path patterns in HINs. 
However, their algorithm does not easily extend to more complicated patterns such as trees; our {\baselinetwo} baseline attempts to adapt their algorithm for ranking tree patterns, and our experiments shows a considerable advantage of {\ouralgorithmname}.

\introparagraph{Top-k query evaluation in databases} 
There is considerable amount of work on top-k queries in a ranking environment~\cite{DBLP:journals/jcss/FaginLN03, DBLP:journals/csur/IlyasBS08, Akbarinia:2007:BPA:1325851.1325909, DBLP:conf/vldb/NatsevCSLV01}. This work aims at minimizing I/O cost by trading sorted access vs.\ random access to data. In contrast to that body of work, we focus on main memory applications and a different cost model.

\introparagraph{Graph search on RDFs and XML}
Top-$k$ keyword search algorithms for XML databases~\cite{XMLtopK10} combine semantic pruning based on document structure encoding with top-$k$ join algorithms from relational databases. The main challenge lies in dealing with query semantics based on least common ancestors. RDF is a flexible and extensible way to represent information about World Wide Web resources. Searching for a pattern on RDFs can be represented in SPARQL, and can be applied to ontology matching~\cite{euzenat2007ontology}.  Recent work on graph pattern matching in RDF databases~\cite[Chapter 2]{A15} has resulted in several different approaches to database layout~\cite[Chapter 3]{A15}. However, as in the case of top-$k$ query evaluation, it appears that more focus has been placed on scalability issues, such as replication, parallelization, and distribution of workloads~\cite{HAR11,HS13}, as RDF datasets are often too large for a single machine. It is well known that SPARQL is descriptive enough to capture graph pattern matching queries (so-called basic graph patterns~\cite[Chapter 2]{A15}), and these queries are typically decomposed into combinations of database primitive operations such as joins, unions, difference, etc.~\cite[Page 23]{A15}. Although work has been done optimizing these primitive operations in the context of graph patterns for certain types of queries that appear in practice~\cite{NW10}, we are not aware of a similar approach to {\ouralgorithmname} being employed for general tree patterns in the context of RDF.

\section{Conclusion}
\label{sec:conclusion}

We proposed {\ouralgorithmname} for finding tree patterns in labeled graphs, e.g., heterogeneous information networks. Compared to previous work, it combines two unique properties. First, it is a top-$k$ anytime algorithm, in the sense that it quickly returns the top-ranked results, then incrementally delivers more on request. This is achieved without sacrificing performance for full-result retrieval compared to bulk-computation. Second, while being subject to the same general hardness of graph isomorphism, {\ouralgorithmname} aggressively exploits the special properties of HINs. We demonstrate this by proving surprisingly strong theoretical guarantees that connect space and time complexity to parameters affected by heterogeneity: result cardinality for the slightly relaxed graph homomorphism problem and number of adjacent edges of a given type. The formulas show that greater heterogeneity of graph and query labels works in our favor by reducing the ``gap'' between homomorphism and isomorphism, and by reducing result size. In future work we will attempt to extend the approach to query patterns with cycles---which appears to be significantly more challenging. Intuitively, it is more challenging to perform elimination of spurious node and edge candidates when pruning based only on local neighborhoods.

\vspace{2mm}
\noindent
\textbf{Acknowledgments}.
This work was supported in part by the National Institutes of Health (NIH) under award number
R01 NS091421 and by the National Science Foundation (NSF) under award number CAREER III-1762268. 
The content is solely the responsibility of the authors and does not necessarily represent the
official views of NIH or NSF. We would also like to thank the reviewers for their constructive feedback.

\clearpage

 \bibliographystyle{ACM-Reference-Format}
\balance
\bibliography{refs} 


\begin{thebibliography}{42}

%%% the Web address.

\ifx \showCODEN    \undefined \def \showCODEN     #1{\unskip}     \fi
\ifx \showDOI      \undefined \def \showDOI       #1{#1}\fi
\ifx \showISBNx    \undefined \def \showISBNx     #1{\unskip}     \fi
\ifx \showISBNxiii \undefined \def \showISBNxiii  #1{\unskip}     \fi
\ifx \showISSN     \undefined \def \showISSN      #1{\unskip}     \fi
\ifx \showLCCN     \undefined \def \showLCCN      #1{\unskip}     \fi
\ifx \shownote     \undefined \def \shownote      #1{#1}          \fi
\ifx \showarticletitle \undefined \def \showarticletitle #1{#1}   \fi
\ifx \showURL      \undefined \def \showURL       {\relax}        \fi
% invisible to TeX
\providecommand\bibfield[2]{#2}
\providecommand\bibinfo[2]{#2}
\providecommand\natexlab[1]{#1}
\providecommand\showeprint[2][]{arXiv:#2}

\bibitem[\protect\citeauthoryear{??}{Yel}{2017}]%
        {Yelp}
 \bibinfo{year}{2017}\natexlab{}.
\newblock \bibinfo{title}{Yelp data set}.
\newblock   (\bibinfo{year}{2017}).
\newblock
\showURL{%
\url{https://www.yelp.com/dataset_challenge/dataset}}


\bibitem[\protect\citeauthoryear{??}{Cod}{2018}]%
        {Code2018}
 \bibinfo{year}{2018}\natexlab{}.
\newblock \bibinfo{title}{Any-k: anytime top-k pattern retrieval in labeled
  graphs (code)}.
\newblock   (\bibinfo{year}{2018}).
\newblock
\showURL{%
\url{https://github.com/northeastern-datalab/Any-k-KARPET}}


\bibitem[\protect\citeauthoryear{??}{Fli}{2018}]%
        {Flicker}
 \bibinfo{year}{2018}\natexlab{}.
\newblock \bibinfo{title}{Flickr}.
\newblock   (\bibinfo{year}{2018}).
\newblock
\showURL{%
\url{http://www.flickr.com/}}


\bibitem[\protect\citeauthoryear{??}{Vit}{2018}]%
        {Vitrage}
 \bibinfo{year}{2018}\natexlab{}.
\newblock \bibinfo{title}{Vitrage}.
\newblock   (\bibinfo{year}{2018}).
\newblock
\showURL{%
\url{https://wiki.openstack.org/wiki/Vitrage}}


\bibitem[\protect\citeauthoryear{Aggarwal and Wang}{Aggarwal and Wang}{2010}]%
        {Aggarwal2010:graphSurvey}
\bibfield{author}{\bibinfo{person}{Charu~C. Aggarwal} {and}
  \bibinfo{person}{Haixun Wang}.} \bibinfo{year}{2010}\natexlab{}.
\newblock \bibinfo{booktitle}{{\em Graph Data Management and Mining: A Survey
  of Algorithms and Applications}}.
\newblock \bibinfo{publisher}{Springer}, \bibinfo{pages}{13--68}.
\newblock


\bibitem[\protect\citeauthoryear{Akbarinia, Pacitti, and Valduriez}{Akbarinia
  et~al\mbox{.}}{2007}]%
        {Akbarinia:2007:BPA:1325851.1325909}
\bibfield{author}{\bibinfo{person}{Reza Akbarinia}, \bibinfo{person}{Esther
  Pacitti}, {and} \bibinfo{person}{Patrick Valduriez}.}
  \bibinfo{year}{2007}\natexlab{}.
\newblock \showarticletitle{Best Position Algorithms for Top-k Queries}. In
  \bibinfo{booktitle}{{\em Proc. VLDB}}. \bibinfo{pages}{495--506}.
\newblock
\showISBNx{978-1-59593-649-3}


\bibitem[\protect\citeauthoryear{Aluc}{Aluc}{2015}]%
        {A15}
\bibfield{author}{\bibinfo{person}{Gunes Aluc}.}
  \bibinfo{year}{2015}\natexlab{}.
\newblock {\em \bibinfo{title}{Workload Matters: {A} Robust Approach to
  Physical {RDF} Database Design}}.
\newblock \bibinfo{thesistype}{Ph.D. Dissertation}. \bibinfo{school}{University
  of Waterloo, Ontario, Canada}.
\newblock


\bibitem[\protect\citeauthoryear{Chen and Papakonstantinou}{Chen and
  Papakonstantinou}{2010}]%
        {XMLtopK10}
\bibfield{author}{\bibinfo{person}{L.~J. Chen} {and} \bibinfo{person}{Y.
  Papakonstantinou}.} \bibinfo{year}{2010}\natexlab{}.
\newblock \showarticletitle{Supporting top-K keyword search in XML databases}.
  In \bibinfo{booktitle}{{\em Proc. ICDE}}. \bibinfo{pages}{689--700}.
\newblock


\bibitem[\protect\citeauthoryear{Cheng, Ke, and Ng}{Cheng
  et~al\mbox{.}}{2009}]%
        {cheng2009efficient}
\bibfield{author}{\bibinfo{person}{James Cheng}, \bibinfo{person}{Yiping Ke},
  {and} \bibinfo{person}{Wilfred Ng}.} \bibinfo{year}{2009}\natexlab{}.
\newblock \showarticletitle{Efficient processing of group-oriented connection
  queries in a large graph}. In \bibinfo{booktitle}{{\em CIKM}}. ACM,
  \bibinfo{pages}{1481--1484}.
\newblock


\bibitem[\protect\citeauthoryear{Cordella, Foggia, Sansone, and Vento}{Cordella
  et~al\mbox{.}}{2001}]%
        {cordella2001improved}
\bibfield{author}{\bibinfo{person}{Luigi~Pietro Cordella},
  \bibinfo{person}{Pasquale Foggia}, \bibinfo{person}{Carlo Sansone}, {and}
  \bibinfo{person}{Mario Vento}.} \bibinfo{year}{2001}\natexlab{}.
\newblock \showarticletitle{An improved algorithm for matching large graphs}.
  In \bibinfo{booktitle}{{\em Proc. IAPR-TC15 Workshop on Graph-based
  Representations in Pattern Recognition}}. \bibinfo{pages}{149--159}.
\newblock


\bibitem[\protect\citeauthoryear{Cordella, Foggia, Sansone, and Vento}{Cordella
  et~al\mbox{.}}{2004}]%
        {cordella2004sub}
\bibfield{author}{\bibinfo{person}{Luigi~P Cordella}, \bibinfo{person}{Pasquale
  Foggia}, \bibinfo{person}{Carlo Sansone}, {and} \bibinfo{person}{Mario
  Vento}.} \bibinfo{year}{2004}\natexlab{}.
\newblock \showarticletitle{A (sub)graph isomorphism algorithm for matching
  large graphs}.
\newblock \bibinfo{journal}{{\em IEEE Trans. on Pattern Analysis and Machine
  Intelligence\/}} \bibinfo{volume}{26}, \bibinfo{number}{10}
  (\bibinfo{year}{2004}), \bibinfo{pages}{1367--1372}.
\newblock


\bibitem[\protect\citeauthoryear{Driscoll, Gabow, Shrairman, and
  Tarjan}{Driscoll et~al\mbox{.}}{1988}]%
        {Driscoll:1988:RHA:50087.50096}
\bibfield{author}{\bibinfo{person}{James~R. Driscoll},
  \bibinfo{person}{Harold~N. Gabow}, \bibinfo{person}{Ruth Shrairman}, {and}
  \bibinfo{person}{Robert~E. Tarjan}.} \bibinfo{year}{1988}\natexlab{}.
\newblock \showarticletitle{Relaxed Heaps: An Alternative to Fibonacci Heaps
  with Applications to Parallel Computation}.
\newblock \bibinfo{journal}{{\em Commun. ACM\/}} \bibinfo{volume}{31},
  \bibinfo{number}{11} (\bibinfo{date}{Nov.} \bibinfo{year}{1988}),
  \bibinfo{pages}{1343--1354}.
\newblock
\showISSN{0001-0782}
\showDOI{%
\url{https://doi.org/10.1145/50087.50096}}


\bibitem[\protect\citeauthoryear{Euzenat, Shvaiko, et~al\mbox{.}}{Euzenat
  et~al\mbox{.}}{2007}]%
        {euzenat2007ontology}
\bibfield{author}{\bibinfo{person}{J{\'e}r{\^o}me Euzenat},
  \bibinfo{person}{Pavel Shvaiko}, {et~al\mbox{.}}}
  \bibinfo{year}{2007}\natexlab{}.
\newblock \bibinfo{booktitle}{{\em Ontology matching}}.
  Vol.~\bibinfo{volume}{18}.
\newblock \bibinfo{publisher}{Springer}.
\newblock


\bibitem[\protect\citeauthoryear{Fagin, Lotem, and Naor}{Fagin
  et~al\mbox{.}}{2003}]%
        {DBLP:journals/jcss/FaginLN03}
\bibfield{author}{\bibinfo{person}{Ronald Fagin}, \bibinfo{person}{Amnon
  Lotem}, {and} \bibinfo{person}{Moni Naor}.} \bibinfo{year}{2003}\natexlab{}.
\newblock \showarticletitle{Optimal aggregation algorithms for middleware}.
\newblock \bibinfo{journal}{{\em J. Comput. Syst. Sci.\/}}
  \bibinfo{volume}{66}, \bibinfo{number}{4} (\bibinfo{year}{2003}),
  \bibinfo{pages}{614--656}.
\newblock


\bibitem[\protect\citeauthoryear{Faloutsos, McCurley, and Tomkins}{Faloutsos
  et~al\mbox{.}}{2004}]%
        {faloutsos2004fast}
\bibfield{author}{\bibinfo{person}{Christos Faloutsos},
  \bibinfo{person}{Kevin~S McCurley}, {and} \bibinfo{person}{Andrew Tomkins}.}
  \bibinfo{year}{2004}\natexlab{}.
\newblock \showarticletitle{Fast discovery of connection subgraphs}. In
  \bibinfo{booktitle}{{\em Proc. ACM SIGKDD}}. \bibinfo{pages}{118--127}.
\newblock


\bibitem[\protect\citeauthoryear{Feng}{Feng}{2014}]%
        {Feng14}
\bibfield{author}{\bibinfo{person}{Gang Feng}.}
  \bibinfo{year}{2014}\natexlab{}.
\newblock \showarticletitle{Finding \emph{k} shortest simple paths in directed
  graphs: {A} node classification algorithm}.
\newblock \bibinfo{journal}{{\em Networks\/}} \bibinfo{volume}{64},
  \bibinfo{number}{1} (\bibinfo{year}{2014}), \bibinfo{pages}{6--17}.
\newblock


\bibitem[\protect\citeauthoryear{Han, Sun, Yan, and Yu}{Han
  et~al\mbox{.}}{2010}]%
        {han2010mining}
\bibfield{author}{\bibinfo{person}{Jiawei Han}, \bibinfo{person}{Yizhou Sun},
  \bibinfo{person}{Xifeng Yan}, {and} \bibinfo{person}{Philip~S Yu}.}
  \bibinfo{year}{2010}\natexlab{}.
\newblock \showarticletitle{Mining knowledge from databases: an information
  network analysis approach}. In \bibinfo{booktitle}{{\em Proc. ACM SIGMOD}}.
  \bibinfo{pages}{1251--1252}.
\newblock


\bibitem[\protect\citeauthoryear{He and Singh}{He and Singh}{2008}]%
        {he2008graphs}
\bibfield{author}{\bibinfo{person}{Huahai He} {and} \bibinfo{person}{Ambuj~K
  Singh}.} \bibinfo{year}{2008}\natexlab{}.
\newblock \showarticletitle{Graphs-at-a-time: query language and access methods
  for graph databases}. In \bibinfo{booktitle}{{\em Proc. ACM SIGMOD}}.
  \bibinfo{pages}{405--418}.
\newblock


\bibitem[\protect\citeauthoryear{Hershberger, Maxel, and Suri}{Hershberger
  et~al\mbox{.}}{2007}]%
        {Hershberger:2007:FKS:1290672.1290682}
\bibfield{author}{\bibinfo{person}{John Hershberger}, \bibinfo{person}{Matthew
  Maxel}, {and} \bibinfo{person}{Subhash Suri}.}
  \bibinfo{year}{2007}\natexlab{}.
\newblock \showarticletitle{Finding the K Shortest Simple Paths: A New
  Algorithm and Its Implementation}.
\newblock \bibinfo{journal}{{\em ACM Trans. Algorithms\/}} \bibinfo{volume}{3},
  \bibinfo{number}{4}, Article \bibinfo{articleno}{45} (\bibinfo{date}{Nov.}
  \bibinfo{year}{2007}).
\newblock
\showISSN{1549-6325}


\bibitem[\protect\citeauthoryear{Hose and Schenkel}{Hose and Schenkel}{2013}]%
        {HS13}
\bibfield{author}{\bibinfo{person}{Katja Hose} {and} \bibinfo{person}{Ralf
  Schenkel}.} \bibinfo{year}{2013}\natexlab{}.
\newblock \showarticletitle{{WARP:} Workload-aware replication and partitioning
  for {RDF}}. In \bibinfo{booktitle}{{\em Proc.\ ICDE Workshops}}.
  \bibinfo{pages}{1--6}.
\newblock


\bibitem[\protect\citeauthoryear{Huang, Abadi, and Ren}{Huang
  et~al\mbox{.}}{2011}]%
        {HAR11}
\bibfield{author}{\bibinfo{person}{Jiewen Huang}, \bibinfo{person}{Daniel~J.
  Abadi}, {and} \bibinfo{person}{Kun Ren}.} \bibinfo{year}{2011}\natexlab{}.
\newblock \showarticletitle{Scalable {SPARQL} Querying of Large {RDF} Graphs}.
\newblock \bibinfo{journal}{{\em PVLDB\/}} \bibinfo{volume}{4},
  \bibinfo{number}{11} (\bibinfo{year}{2011}), \bibinfo{pages}{1123--1134}.
\newblock


\bibitem[\protect\citeauthoryear{Ilyas, Beskales, and Soliman}{Ilyas
  et~al\mbox{.}}{2008}]%
        {DBLP:journals/csur/IlyasBS08}
\bibfield{author}{\bibinfo{person}{Ihab~F. Ilyas}, \bibinfo{person}{George
  Beskales}, {and} \bibinfo{person}{Mohamed~A. Soliman}.}
  \bibinfo{year}{2008}\natexlab{}.
\newblock \showarticletitle{A survey of top-\emph{k} query processing
  techniques in relational database systems}.
\newblock \bibinfo{journal}{{\em {ACM} Comput. Surv.\/}} \bibinfo{volume}{40},
  \bibinfo{number}{4} (\bibinfo{year}{2008}), \bibinfo{pages}{11:1--11:58}.
\newblock


\bibitem[\protect\citeauthoryear{Kasneci, Elbassuoni, and Weikum}{Kasneci
  et~al\mbox{.}}{2009}]%
        {kasneci2009ming}
\bibfield{author}{\bibinfo{person}{Gjergji Kasneci}, \bibinfo{person}{Shady
  Elbassuoni}, {and} \bibinfo{person}{Gerhard Weikum}.}
  \bibinfo{year}{2009}\natexlab{}.
\newblock \showarticletitle{Ming: mining informative entity relationship
  subgraphs}. In \bibinfo{booktitle}{{\em Proc. CIKM}}. ACM,
  \bibinfo{pages}{1653--1656}.
\newblock


\bibitem[\protect\citeauthoryear{Katoh, Ibaraki, and Mine}{Katoh
  et~al\mbox{.}}{1982}]%
        {katoh}
\bibfield{author}{\bibinfo{person}{Naoki Katoh}, \bibinfo{person}{Toshihide
  Ibaraki}, {and} \bibinfo{person}{Hisashi Mine}.}
  \bibinfo{year}{1982}\natexlab{}.
\newblock \showarticletitle{An efficient algorithm for k shortest simple
  paths}.
\newblock \bibinfo{journal}{{\em Networks\/}} \bibinfo{volume}{12},
  \bibinfo{number}{4} (\bibinfo{year}{1982}), \bibinfo{pages}{411--427}.
\newblock


\bibitem[\protect\citeauthoryear{Koutis and Williams}{Koutis and
  Williams}{2016}]%
        {KW16}
\bibfield{author}{\bibinfo{person}{Ioannis Koutis} {and} \bibinfo{person}{Ryan
  Williams}.} \bibinfo{year}{2016}\natexlab{}.
\newblock \showarticletitle{{LIMITS} and Applications of Group Algebras for
  Parameterized Problems}.
\newblock \bibinfo{journal}{{\em {ACM} Trans. Algorithms\/}}
  \bibinfo{volume}{12}, \bibinfo{number}{3} (\bibinfo{year}{2016}),
  \bibinfo{pages}{31:1--31:18}.
\newblock


\bibitem[\protect\citeauthoryear{Krauthgamer and Trabelsi}{Krauthgamer and
  Trabelsi}{2017}]%
        {KT17}
\bibfield{author}{\bibinfo{person}{Robert Krauthgamer} {and}
  \bibinfo{person}{Ohad Trabelsi}.} \bibinfo{year}{2017}\natexlab{}.
\newblock \bibinfo{booktitle}{{\em Conditional Lower Bound for Subgraph
  Isomorphism with a Tree Pattern}}.
\newblock \bibinfo{type}{{T}echnical {R}eport}.
\newblock
\showURL{%
\url{https://arxiv.org/abs/1708.07591}}


\bibitem[\protect\citeauthoryear{Lee, Han, Kasperovics, and Lee}{Lee
  et~al\mbox{.}}{2012}]%
        {lee2012depth}
\bibfield{author}{\bibinfo{person}{Jinsoo Lee}, \bibinfo{person}{Wook-Shin
  Han}, \bibinfo{person}{Romans Kasperovics}, {and} \bibinfo{person}{Jeong-Hoon
  Lee}.} \bibinfo{year}{2012}\natexlab{}.
\newblock \showarticletitle{An In-depth Comparison of Subgraph Isomorphism
  Algorithms in Graph Databases}.
\newblock \bibinfo{journal}{{\em PVLDB\/}} \bibinfo{volume}{6},
  \bibinfo{number}{2} (\bibinfo{year}{2012}), \bibinfo{pages}{133--144}.
\newblock


\bibitem[\protect\citeauthoryear{Liang, Ajwani, Nicholson, Sala, and
  Parthasarathy}{Liang et~al\mbox{.}}{2016}]%
        {liang2016links}
\bibfield{author}{\bibinfo{person}{Jiongqian Liang}, \bibinfo{person}{Deepak
  Ajwani}, \bibinfo{person}{Patrick~K Nicholson}, \bibinfo{person}{Alessandra
  Sala}, {and} \bibinfo{person}{Srinivasan Parthasarathy}.}
  \bibinfo{year}{2016}\natexlab{}.
\newblock \showarticletitle{What Links Alice and Bob? Matching and Ranking
  Semantic Patterns in Heterogeneous Networks}. In \bibinfo{booktitle}{{\em
  Proc. WWW}}. \bibinfo{pages}{879--889}.
\newblock


\bibitem[\protect\citeauthoryear{Lubiw}{Lubiw}{1981}]%
        {lubiw1981some}
\bibfield{author}{\bibinfo{person}{Anna Lubiw}.}
  \bibinfo{year}{1981}\natexlab{}.
\newblock \showarticletitle{Some {NP}-complete problems similar to graph
  isomorphism}.
\newblock \bibinfo{journal}{{\it SIAM J. Comput.}} \bibinfo{volume}{10},
  \bibinfo{number}{1} (\bibinfo{year}{1981}), \bibinfo{pages}{11--21}.
\newblock


\bibitem[\protect\citeauthoryear{Natsev, Chang, Smith, Li, and Vitter}{Natsev
  et~al\mbox{.}}{2001}]%
        {DBLP:conf/vldb/NatsevCSLV01}
\bibfield{author}{\bibinfo{person}{Apostol Natsev}, \bibinfo{person}{Yuan{-}Chi
  Chang}, \bibinfo{person}{John~R. Smith}, \bibinfo{person}{Chung{-}Sheng Li},
  {and} \bibinfo{person}{Jeffrey~Scott Vitter}.}
  \bibinfo{year}{2001}\natexlab{}.
\newblock \showarticletitle{Supporting Incremental Join Queries on Ranked
  Inputs}. In \bibinfo{booktitle}{{\em Proc.\ VLDB}}.
  \bibinfo{pages}{281--290}.
\newblock


\bibitem[\protect\citeauthoryear{Neumann and Weikum}{Neumann and
  Weikum}{2010}]%
        {NW10}
\bibfield{author}{\bibinfo{person}{Thomas Neumann} {and}
  \bibinfo{person}{Gerhard Weikum}.} \bibinfo{year}{2010}\natexlab{}.
\newblock \showarticletitle{The {RDF-3X} engine for scalable management of
  {RDF} data}.
\newblock \bibinfo{journal}{{\em The VLDB Journal\/}} \bibinfo{volume}{19},
  \bibinfo{number}{1} (\bibinfo{year}{2010}), \bibinfo{pages}{91--113}.
\newblock


\bibitem[\protect\citeauthoryear{Ramakrishnan, Milnor, Perry, and
  Sheth}{Ramakrishnan et~al\mbox{.}}{2005}]%
        {ramakrishnan2005discovering}
\bibfield{author}{\bibinfo{person}{Cartic Ramakrishnan},
  \bibinfo{person}{William~H Milnor}, \bibinfo{person}{Matthew Perry}, {and}
  \bibinfo{person}{Amit~P Sheth}.} \bibinfo{year}{2005}\natexlab{}.
\newblock \showarticletitle{Discovering informative connection subgraphs in
  multi-relational graphs}.
\newblock \bibinfo{journal}{{\em ACM SIGKDD Explorations Newsletter\/}}
  \bibinfo{volume}{7}, \bibinfo{number}{2} (\bibinfo{year}{2005}),
  \bibinfo{pages}{56--63}.
\newblock


\bibitem[\protect\citeauthoryear{Roditty}{Roditty}{2007}]%
        {Roditty07}
\bibfield{author}{\bibinfo{person}{Liam Roditty}.}
  \bibinfo{year}{2007}\natexlab{}.
\newblock \showarticletitle{On the \emph{K}-simple shortest paths problem in
  weighted directed graphs}. In \bibinfo{booktitle}{{\em Proc. SODA}}.
  \bibinfo{pages}{920--928}.
\newblock


\bibitem[\protect\citeauthoryear{Shang, Zhang, Lin, and Yu}{Shang
  et~al\mbox{.}}{2008}]%
        {shang2008taming}
\bibfield{author}{\bibinfo{person}{Haichuan Shang}, \bibinfo{person}{Ying
  Zhang}, \bibinfo{person}{Xuemin Lin}, {and} \bibinfo{person}{Jeffrey~Xu Yu}.}
  \bibinfo{year}{2008}\natexlab{}.
\newblock \showarticletitle{Taming Verification Hardness: An Efficient
  Algorithm for Testing Subgraph Isomorphism}.
\newblock \bibinfo{journal}{{\em PVLDB\/}} \bibinfo{volume}{1},
  \bibinfo{number}{1} (\bibinfo{year}{2008}), \bibinfo{pages}{364--375}.
\newblock


\bibitem[\protect\citeauthoryear{Sun, Han, Yan, Yu, and Wu}{Sun
  et~al\mbox{.}}{2011}]%
        {sun2011pathsim}
\bibfield{author}{\bibinfo{person}{Yizhou Sun}, \bibinfo{person}{Jiawei Han},
  \bibinfo{person}{Xifeng Yan}, \bibinfo{person}{Philip~S Yu}, {and}
  \bibinfo{person}{Tianyi Wu}.} \bibinfo{year}{2011}\natexlab{}.
\newblock \showarticletitle{Pathsim: Meta path-based top-k similarity search in
  heterogeneous information networks}.
\newblock \bibinfo{journal}{{\em PVLDB\/}} \bibinfo{volume}{4},
  \bibinfo{number}{11} (\bibinfo{year}{2011}), \bibinfo{pages}{992--1003}.
\newblock


\bibitem[\protect\citeauthoryear{Tong and Faloutsos}{Tong and
  Faloutsos}{2006}]%
        {tong2006center}
\bibfield{author}{\bibinfo{person}{Hanghang Tong} {and}
  \bibinfo{person}{Christos Faloutsos}.} \bibinfo{year}{2006}\natexlab{}.
\newblock \showarticletitle{Center-piece subgraphs: problem definition and fast
  solutions}. In \bibinfo{booktitle}{{\em Proc. ACM SIGKDD}}.
  \bibinfo{pages}{404--413}.
\newblock


\bibitem[\protect\citeauthoryear{Ullmann}{Ullmann}{1976}]%
        {ullmann1976algorithm}
\bibfield{author}{\bibinfo{person}{Julian~R Ullmann}.}
  \bibinfo{year}{1976}\natexlab{}.
\newblock \showarticletitle{An algorithm for subgraph isomorphism}.
\newblock \bibinfo{journal}{{\em {JACM}\/}} \bibinfo{volume}{23},
  \bibinfo{number}{1} (\bibinfo{year}{1976}), \bibinfo{pages}{31--42}.
\newblock


\bibitem[\protect\citeauthoryear{Wei}{Wei}{2010}]%
        {wei2010efficient}
\bibfield{author}{\bibinfo{person}{Fang Wei}.} \bibinfo{year}{2010}\natexlab{}.
\newblock \showarticletitle{Efficient graph reachability query answering using
  tree decomposition}. In \bibinfo{booktitle}{{\em Int. Workshop on
  Reachability Problems}}. \bibinfo{pages}{183--197}.
\newblock


\bibitem[\protect\citeauthoryear{Yang, Ajwani, Gatterbauer, Nicholson,
  Riedewald, and Sala}{Yang et~al\mbox{.}}{2018}]%
        {yang2018}
\bibfield{author}{\bibinfo{person}{Xiaofeng Yang}, \bibinfo{person}{Deepak
  Ajwani}, \bibinfo{person}{Wolfgang Gatterbauer}, \bibinfo{person}{Patrick~K
  Nicholson}, \bibinfo{person}{Mirek Riedewald}, {and}
  \bibinfo{person}{Alessandra Sala}.} \bibinfo{year}{2018}\natexlab{}.
\newblock \bibinfo{booktitle}{{\em Any-k: anytime top-k pattern retrieval in
  labeled graphs}}.
\newblock \bibinfo{type}{{T}echnical {R}eport}.
\newblock
\showURL{%
\url{https://arxiv.org/abs/1802.06060}}


\bibitem[\protect\citeauthoryear{Yannakakis}{Yannakakis}{1981}]%
        {DBLP:conf/vldb/Yannakakis81}
\bibfield{author}{\bibinfo{person}{Mihalis Yannakakis}.}
  \bibinfo{year}{1981}\natexlab{}.
\newblock \showarticletitle{Algorithms for Acyclic Database Schemes}. In
  \bibinfo{booktitle}{{\em Proc. VLDB}}. \bibinfo{pages}{82--94}.
\newblock


\bibitem[\protect\citeauthoryear{Yen}{Yen}{1971}]%
        {doi:10.1287/mnsc.17.11.712}
\bibfield{author}{\bibinfo{person}{Jin~Y. Yen}.}
  \bibinfo{year}{1971}\natexlab{}.
\newblock \showarticletitle{Finding the K Shortest Loopless Paths in a
  Network}.
\newblock \bibinfo{journal}{{\em Management Science\/}} \bibinfo{volume}{17},
  \bibinfo{number}{11} (\bibinfo{year}{1971}), \bibinfo{pages}{712--716}.
\newblock


\bibitem[\protect\citeauthoryear{Zhang, Li, and Yang}{Zhang
  et~al\mbox{.}}{2009}]%
        {zhang2009gaddi}
\bibfield{author}{\bibinfo{person}{Shijie Zhang}, \bibinfo{person}{Shirong Li},
  {and} \bibinfo{person}{Jiong Yang}.} \bibinfo{year}{2009}\natexlab{}.
\newblock \showarticletitle{GADDI: distance index based subgraph matching in
  biological networks}. In \bibinfo{booktitle}{{\em Proc. EDBT}}.
  \bibinfo{pages}{192--203}.
\newblock


\end{thebibliography}

\clearpage
\appendix
\section{Appendix: Proof of theorems}

\begin{proof}[Proof \autoref{thm:spuriousNode}]
We sketch the main ideas of the proof. (1) By definition of Alg.~\ref{alg:prioritized-search}, for $c$ to be reached during the top-down search, there has to exist a path $c_1, c_2,\ldots,c_j$, where $c_j=c$, from a root node candidate $c_1$ down to $c$. For each $i$, let $q_i \in V_Q$ denote the query node for which $c_i \in \Candidates(q_i)$ is a candidate. Without loss of generality, let $c_i$ be the \emph{left-most} child of $c_{i-1}$, $i=2,\ldots,j$. We next show that we can construct a query result from the path, the subtree rooted at $c_j$, and the \emph{right} subtrees for all $c_i$, $i < j$.

By definition, Alg.~\ref{alg:pruning} only adds a candidate $c_i$ to a query node $q_i$, if there exists a subtree rooted at $c_i$ in $\Graph$ that is isomorphic to the subtree rooted at $q_i$ in query $\TreePattern$. (Otherwise $c_i$ would not have been reached during the bottom-up edge traversal from \emph{all} its leaves.) It is now easy to see that ($i$) the path $q_1, q_2,\ldots,q_j$, ($ii$) all \emph{right} subtrees of the nodes in $\{q_1, q_2,\ldots, q_{j-1}\}$, and ($iii$) the subtree rooted at $c_j$, together cover the entire query graph and are pairwise disjoint. Hence the corresponding candidate instances, i.e., ($i$) path $c_1, c_2,\ldots,c_j$, ($ii$) the right subtree matches rooted at each candidate in $c_1, c_2,\ldots,c_{j-1}$, and ($iii$) the subtree match rooted at $c_j$, form a query result containing candidate $c_j = c$ for query node $q_j = q$.
\end{proof}

\begin{proof}[Proof \autoref{lem:priorityMonotonicity}]
Recall that the priority of a partial pattern $P$ is defined as the sum of the weights of all matched edges, plus the sum of the pre-computed weights of all unmatched subtrees rooted at the corresponding matched candidates in $P$. 
A candidate's subtree weight, by definition, is the minimum over all possible matches in the subtree rooted at this candidate. Hence any successor of $P$, which might not expand along the edges that determined the minimum subtree weights, will have a priority value greater than or equal to the priority value of $P$.
\end{proof}

\begin{proof}[Proof \autoref{lem:sameWeightPQ}]
Similar to the proof for Lemma~\ref{lem:priorityMonotonicity}, we start with the fact that the priority of a partial pattern $P$ is defined as the sum of the weights of all matched edges, plus the sum of the pre-computed weights of all unmatched subtrees rooted at the corresponding matched candidates in $P$. Since a candidate's subtree weight, by definition, is the minimum over all possible matches in the subtree rooted at this candidate, there exists a full match whose weight is equal to that minimum. For a partial match $P=(c_1, c_2,\ldots, c_i)$, successor $(c_1, c_2,\ldots, c_i, c_{i+1})$ has the same priority $w$ if the edge connecting $c_{i+1}$ to the pattern is the one that determined the minimum weight value for the corresponding subtree in the parent of $c_{i+1}$.
\end{proof}

\begin{proof}[Proof \autoref{thm:kPop}]
Consider two results $P=(c_1, c_2,\ldots, c_{|V_Q|})$ and $P'=(c'_1, c'_2,\ldots, c_{|V_Q|})$ with weights $w$ and $w'$, respectively, such that $w < w'$. Assume that the prefix $\pi$ of $P$ that was expanded into full result $P$ was popped at time $j$; and similarly for prefix $\pi'$ of $P'$ and time $j'$. We now prove that $j < j'$. Lemma~\ref{lem:sameWeightPQ} implies that the priority of $\pi'$ at time $j'$ is $w'$. Similarly, Lemmas~\ref{lem:priorityMonotonicity} and \ref{lem:sameWeightPQ} imply that at time $j'$, no prefix of $P$ could have been in \texttt{pq}. Any such prefix of $P$ has priority less than or equal to $w$, hence would have been in front of $\pi'$ at time $j'$. By definition, Alg.~\ref{alg:prioritized-search} can only push successors of a pattern it pops from \texttt{pq}. This implies that it can only push patterns that are successors of those partial matches currently in \texttt{pq}. (Note that it initially pushes all root candidates, guaranteeing the 1-node prefixes of all result patterns are initially in \texttt{pq}.) Since at time $j'$ there is no prefix of $P$ in \texttt{pq}, Alg.~\ref{alg:prioritized-search} cannot push or pop any such prefix after time $j'$. This in turn implies that it must have popped $\pi$ in an earlier pop operation before $j'$.
\end{proof}

\begin{proof}[Proof \autoref{thm:spaceAlg2}]
The upper bound follows from Corollary~\ref{cor:numPushes}.
We show that the bound is tight: Assume all results have the same prefix $(c_1,\ldots, c_{|V_Q|-1})$, except for the last node. Then Alg.~\ref{alg:prioritized-search} pushes only root $c_1$, immediately pops it, then expands until reaching $(c_1,\ldots, c_{|V_Q|-1})$. The next expansion produces \#results many patterns that are all pushed to \texttt{pq}, except for the minimal one. Hence at this point there are (\#results - 1) elements in \texttt{pq}, each consisting of $|V_Q|$ matched nodes.
\end{proof}

\begin{proof}[Proof \autoref{thm:interarrival}]

Theorem~\ref{thm:spuriousNode} implies $\mathrm{outdegree} \le r_H$. Assuming that \texttt{pq} is implemented using a relaxed heap, then it will have worst case time bound of O(1) for \textit{decrease key} and O(log n) for \textit{delete min}\cite{Driscoll:1988:RHA:50087.50096}.  Due to front-element optimization, the next full match is computed by (1) popping the first element from \texttt{pq} in time logarithmic in queue size, which is at most $r_H-1$, then (2) expanding the partial match node-by-node until a full match. The latter requires retrieving all edges from a currently matched node in the partial match to the next unmatched node according to pre-order traversal. There are at most $\mathrm{outDegree}$ such edges, resulting in the corresponding number of pushes to \texttt{pq}, each at (expected) constant cost because of the constant time inserting into a relaxed heap. This is done at most $|\EdgesPattern|$ times, which is a constant (see above).
\hide{This bound is tight. Consider a path query where each node on the shortest path has $d$ children, all others have one. Minor caveat: filling up the pq of size $r_H$ is slightly faster than $r_H \log r_H$.}
\end{proof}

\begin{proof}[Proof \autoref{thm:timeAlg2}]
For the lower bound, note that each of the $r_H$ results has to be output. For the total cost of Alg.~\ref{alg:prioritized-search}, Theorem~\ref{thm:kPop} and Corollary~\ref{cor:numPushes} guarantee that the total number of push and pop operations on \texttt{pq} is $r_H$. Since the former has constant cost, and the latter is logarithmic in queue size, we obtain an upper bound of $\mathrm{O}(r_H \log r_H)$ for operations on \texttt{pq}. (To be precise, for both bulk-computation and Alg.~\ref{alg:prioritized-search}, there is another additive term to account for the number of edges accessed to assemble each result pattern. This number is linear in $r_H$.)
\end{proof}

\end{document}